## Title

Variable bandwidth, high efficiency microwave resonator
for control of spin-qubits in nitrogen-vacancy centers

## Authors

Dr. Anton Savitsky, Dr. Jingfu Zhang, Prof. Dieter Suter

Faculty of Physics, Technical University Dortmund, Otto-Hahn-Str. 4a, 44227 Dortmund, Germany

## Keywords

Planar microresonator, NV center, spin qubit, microwave magnetic field

## Abstract

Nitrogen-Vacancy (NV) centers in diamond are attractive tools for sensing and quantum information. Realisation of this potential requires effective tools for controlling the spin degree of freedom by microwave (mw) magnetic fields. In this work we present a planar microwave resonator optimised for microwave-optical double resonance experiments on single nitrogen-vacancy (NV) centers in diamond. It consists of a piece of wide microstrip line which is symmetrically connected to two 50 Ω microstrip feed lines. In the center of the resonator, an Ω-shaped loop focuses the current and the mw magnetic field. It generates a relatively homogeneous magnetic field over a volume of $0.07\text{mm}^2\times0.1\text{mm}$. It can be operated at 2.9 GHz in both transmission and reflection modes with bandwidths of 1000 MHz and 400 MHz, respectively. The high power-to-magnetic field conversion efficiency allows to produce π-pulses with a duration of 50 ns with only about 200 mW and 50 mW microwave power in transmission and reflection, respectively. The transmission mode also offers capability for efficient radio frequency excitation. The resonance frequency can be tuned between 1.3 GHz and 6 GHz by adjusting the length of the resonator. This will be useful for experiments on NV-centers at higher external magnetic fields and on different types of optically active spin centres.

## 1. Introduction

The NV center in diamond is used in various fields, such as quantum information, quantum sensing, magnetometry, bioimaging, etc. [1-5]. In all applications, efficient manipulation of the electron spin is an essential prerequisite, both for continuous wave (cw) experiments and for fast spin control in pulsed experiments. Therefore, maximising the coupling between the control microwave (mw) magnetic field and the electron spins is a fundamental concern. Here, we consider specifically applications where single NV centers are excited by a laser and the spin state is read out by collecting photoluminescence (PL) through a microscope objective with large numerical aperture (NA > 1). Measuring a cw ODMR spectrum in such requires a mw magnetic field of $H = 150$ A/m ($B = \mu_0 H = 0.19$ mT) to reach the maximum fluorescence contrast [6]. This field corresponds to a Rabi frequency of $f_R = 2$ MHz if the oscillating magnetic field is perpendicular to the NV axis. In pulsed experiments, typical π-pulse durations of $t_\pi = 50$ ns are used, which correspond to an excitation bandwidth of $1.2/t_\pi = 24$ MHz ($f_R = 10$ MHz). This is sufficient for complete electron spin flip of the selected triplet spin transition [6]. This pulse length requires magnetic field amplitude of $H = 800$ A/m.

Currently, most experimental setups rely on a wire positioned over the sample to achieve such mw magnetic field amplitudes. If this wire is placed in a short gap of a transmission line with $Z_L$ =50 Ω impedance, the electric current in the wire has an amplitude $I = \sqrt{2P_{in}/Z_L} =$ $0.2\ \mathrm{A}/\sqrt{\mathrm{W}} \cdot \sqrt{P_{in}}$. This current produces the tangential magnetic field $H_{wire} = \frac{I}{2\pi r}$ outside the wire of radius $r_0$. For a typical wire diameter of 20 μm, the maximum magnetic field generated at the wire surface is $H_{wire}(r_0) = 3200 \frac{A}{m}$ at 1W of mw power. The wire is, however, optically opaque. Therefore, only NV-centers at a distance > $\sqrt{2}r_0$ from the center of the wire center can be optically probed. At this position, the magnetic field is reduced to $2600\ \frac{A}{m}$ at 1W. Even at realistic distances from the wire of 40 μm, which still reduces the fluorescence collection efficiency due to obstruction by the wire, pulsed experiments can be performed with 1 W mw power, reaching $H_{wire}(40\ \mu\mathrm{m}) = 800\ \frac{A}{m}$. Therefore, the microwire system is used in many laboratories, including our laboratory[7]. A major disadvantage is that the $1/r$ dependence of the magnetic field amplitude significantly restricts the volume and the surface area available for probing single NV-centers and requires precise initial positioning of the microscope objective. Moreover, readjustment of the mw power settings is necessary for any new probed NV-center which is not only due to the distance dependence of the magnetic field amplitude but also to the magnetic field direction dependence on distance. This problem can be overcome using a microloop integrated into a transmission line instead of the microwire [8, 9]. The magnetic field amplitude in the center of the ideal loop is given by $H_{loop} = \frac{I}{D}$ where $D$ is the loop diameter. Thus, the field of 1000 $\frac{A}{m}$ can be generated by a loop with 200 μm diameter at 1 W mw power, which fulfills the requirements for ODMR. Advantages of the loop include a good homogeneity and directivity of the magnetic field over a 100 ×100 μm$^2$ area, much higher than the microwire. Compared to a wire, the loop magnetic field is, however, less tolerant to the effect of the metal case of the microscope objective, which must be positioned close to the sample. We discuss means to avoid this issue below.

There are several additional significant handicaps of loop, as well as wire, systems in conjunction with transmission line, *(i)* the construction does not allow quick replacement or reposition of the diamond spaceman; *(ii)* significant heating effects (the resistance of 10 mm copper wire with 20 μm diameter is 0.25 Ω at 3 GHz, which leads to power dissipation at the mw power level required for ODMR, but the thermal conductivity is limited due to the small conductor cross-section); *(iii)* substantial power return losses caused by the discontinuity in the transmission line. The main handicap is, however, the high mw powers required to fulfill the magnetic field requirements for ODMR experiments.

The power limitation can be overcome using planar resonators to generate the magnetic field. In the past numerous reflection mode resonators were proposed based on different planar structures, for instance double-split ring [10, 11], triple-split-ring [12], loop-gap [13], strip-line [14], and several other types [15]. All these resonators allow to store the mw energy for a time proportional to the unloaded quality factor $Q_0$, which leads to increase of the current and, therefore, an enhanced magnetic field amplitude proportional to $\sqrt{Q_0}$ as compared to a non-resonant structure with the same geometry. The resonator, however, introduces additional limitation. The substantial magnetic field enhancement is achieved only within the resonator bandwidth. For instance, for a matched (critically coupled) reflection resonator with $Q_0$=100 would require 100 times less mw power at the resonance frequency of $\nu_0 = 3$ GHz to generate the same mw magnetic field amplitude as compared to the non-resonant structure of same geometry. This effect is, however, only obtained within $\Delta\nu_{1/2} = 2 \cdot Q_0^{-1} \cdot \nu_0$= 60 MHz around the resonance frequency. This would substantially limit its general applicability for ODMR on single NV-centers. The bandwidth of a matched resonator can only be increased by lowering the $Q_0$-value [13]. This, however, leads to lower mw power to magnetic field conversion, i.e. it reduces the usefulness of the resonator.

The aim of this work is to develop a device that avoids these issues and can be used, e.g., for ODMR spectroscopy of single NV centers. It overcomes high power requirements and compromises limitations of transmission line systems as well as previously reported resonant structures. We design it to fulfill the following requirements: *(i)* easy and cheap fabrication with reliable resonator parameters; *(ii)* compatibility with standard coaxial mw delivery system; *(iii)* high mw power to magnetic field conversion efficiency; *(iv)* high magnetic field directivity and spatial homogeneity at least in the area accessible by nanopositioners based on piezoelectric actuator (100 ×100 μm$^2$); *(v)* small mw magnetic field frequency dependence, i.e. large bandwidth; *(vi)* possibility for microwave and radio frequency excitations; *(vii)* high mechanical and thermal stability; (*viii*) easy and reliable replacement or reposition of the diamond specimen; and (very important !) *(ix)* it must be compatible with high resolution confocal objectives, i.e. tolerate the presence of dielectric and conductive parts in very close vicinity (200 μm to 300 μm) of the magnetic fields of the resonator.

## 2. Results and discussion

### 2.1 Resonator design

Figure 1 depicts the design of the developed half-wave resonator. It consists of a piece of wide microstrip line symmetrically connected to two 50 Ω microstrip feed lines terminated by SMA connectors. In the center of the microstrip, the Ω-loop concentrates the current and the mw magnetic field. The geometrical parameters of the resonator investigated in this study are summarized in the figure caption. The diamond is placed on the loop, as shown in Fig 1b. Thus, the lower diamond surface is exposed to the mw magnetic field near the maximum amplitude. The resonator is fabricated using standard PCB lithography on low loss Rogers RO3003 laminate with 166 μm overall thickness. This design allows us to explore the diamond down to least 100 μm above the surface. The resonator is designed for operation in transmission (the output SMA is terminated by a 50 Ω load) or reflection (the output SMA is open to free space) modes of operation.

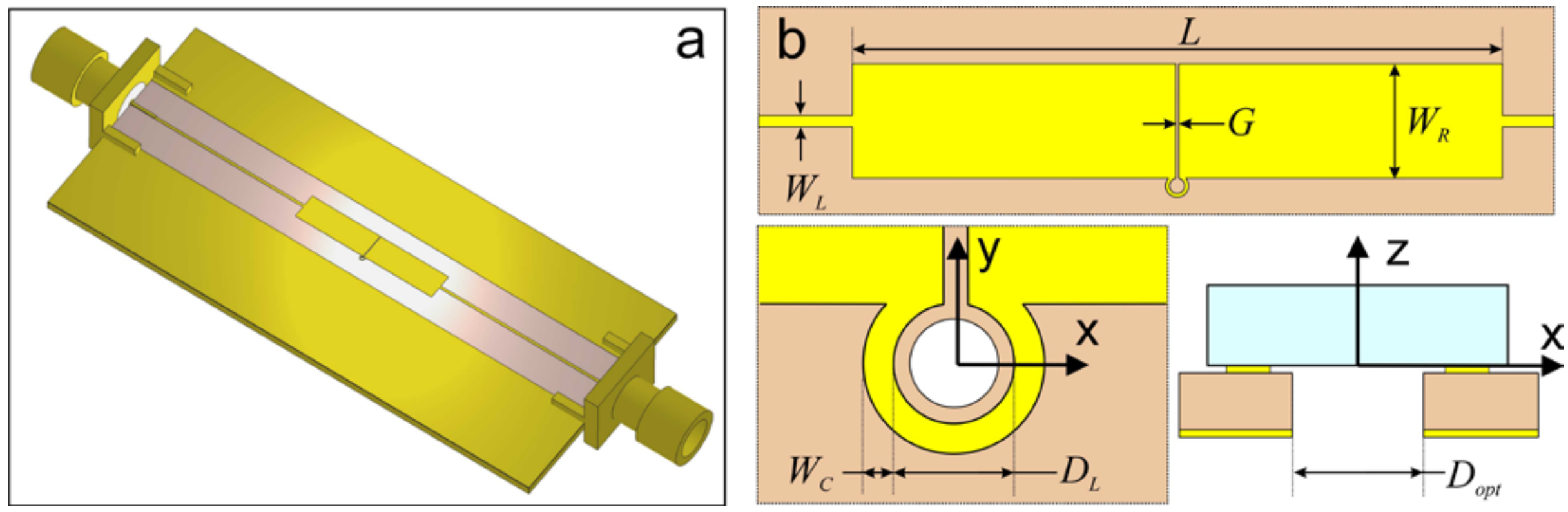

**Figure 1.** (**a**) Full model of the microwave resonator. It is fabricated on 60×26 mm$^2$ Rogers RO3003 low loss laminate ( $\varepsilon_r = 3.00$ , $\tan\delta = 0.001$ at 10 GHz, dielectric thickness 0.13 mm and 18 μm copper cladding) mounted on a 60×26×1mm$^3$ copper holder and two standard PCB SMA flat tab connectors . (**b**) Resonator design. The geometrical parameters are: $W_L = 0.3$ mm - width of the microstrip feed line ($Z_L = 50\ \Omega$); $W_R = 3$ mm - width of the resonator, $L = 17$ mm - length of the resonator; $G = 0.1$ mm - gap width, $D_L = 0.4$ mm - inner diameter of the loop, $W_C = 0.1$ mm - width of the loop conductor, $D_{opt} = 0.3$ mm - diameter of the optical access hole. Coordinate system is indicated. For optical pathway see Fig. S8 in SI.

### 2.2 Transmission mode resonator

Figure 2 shows the S-parameters of the resonator in transmission mode calculated using CST Microwave Studio and measured experimentally using a network vector analyzer (HP 8720A). The analytical transmission and reflection coefficients of the symmetrically coupled transmission resonator with unloaded quality factor $Q_0$ and resonance frequency $\nu_0$ are given by

$$\Gamma = -\frac{1-i\xi}{1+2\beta-i\xi};\ T = \frac{2\beta}{1+2\beta-i\xi} \tag{1}$$

where $\beta$ is the coupling parameter for both input and output and $\xi = Q_0 \cdot \left(\frac{\nu_0}{\nu} - \frac{\nu}{\nu_0}\right)$ is the normalized offset [16, 17]. Analysis of the S-parameter traces depicted in Fig. 2 using Eqs. (1)

yields the resonator parameters $Q_0 = 74$, $\beta = 11.5$ and $Q_0 = 73$, $\beta = 11.8$ for simulation and experiment, respectively. The relatively low unloaded quality factor is typical for microstrip based resonators owing to relatively high conduction losses [18]. The comparison of the calculated coupling parameter of 11.5 with $\beta = Q_0 \frac{Z_R}{Z_L}$=12.6 which can estimated analytically for the resonator shows that the Ω-loop slightly increases the impedance of the resonator over the impedance of the microstrip line with width $W_R$ (for more information see SI). The central frequency of the experimental resonator was downshifted by about 70 MHz compared to the calculation. This deviation is ascribed to slightly different real parameters of the laminate (dielectric constant and thickness) as well as manufacturing tolerances.

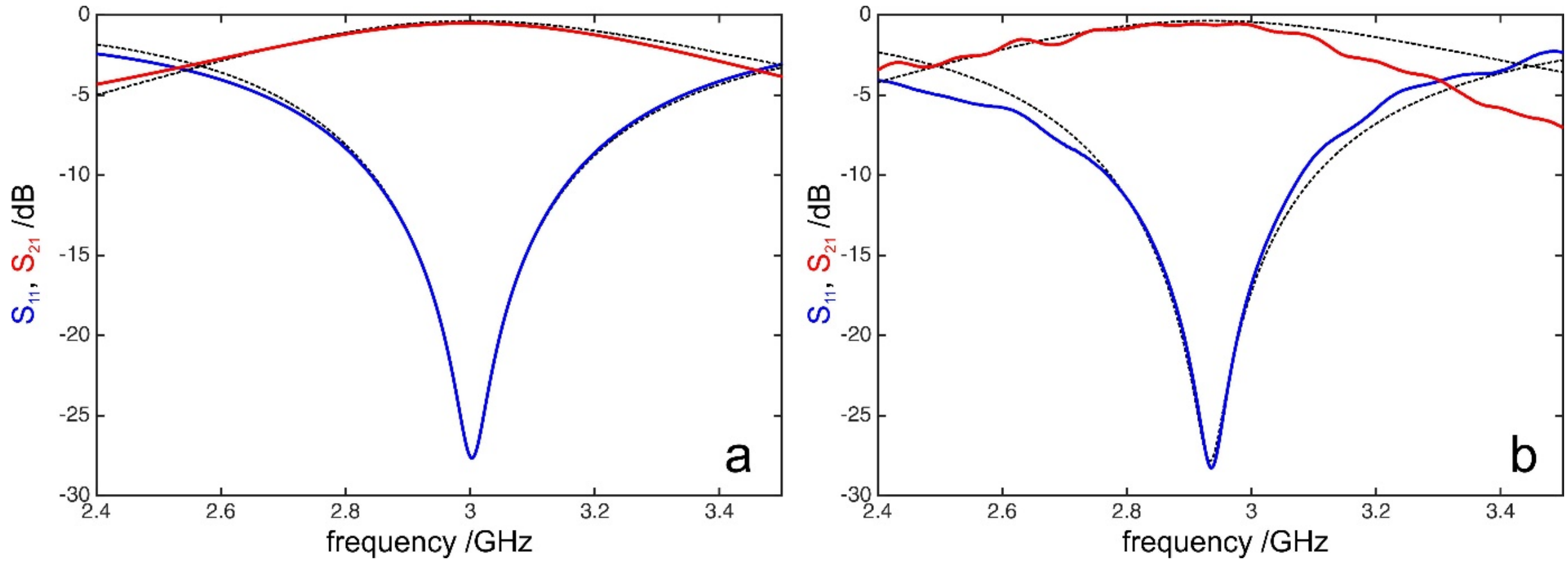

**Figure 2.** **(a)** Simulated and **(b)** experimental S-parameters of the transmission resonator with diamond: $S_{11} = 20 \cdot \log_{10} |\Gamma|$ (blue line) and $S_{21} = 20 \cdot \log_{10} |T|$ (red line). The dashed lines show the best fit results of the S-curves to the reflection and transmission coefficients given by Eqs. (1). The disagreement at the low and high frequencies is due to the frequency dependence of the characteristic impedances, which are not included in Eqs. (1), and contributions from other resonator modes.

The resonator itself is described by the coefficient $R^2 = 1 - |\Gamma|^2 - |T|^2$, i.e. the mw power transmitted to the resonator to which the magnetic field amplitude is proportional. This factor can be parametrized by the power coefficient at the resonance frequency and the resonator bandwidth:

$$R(\nu_0) = \frac{2\sqrt{\beta}}{1+2\beta};\ \Delta\nu_{1/2} = \frac{1+2\beta}{Q_0} \cdot \nu_0. \qquad (2)$$

Thus, the bandwidth of the resonator is 950 MHz and 990 MHz for calculation and experiment, respectively, which is sufficient for most low-field applications of NV centers.

Figure 3 shows the spatial variation of the mw magnetic field behavior within the loop. At a small elevation above the loop surface ($z$ =10 μm), the loop provides a perfect magnetic field directivity over the optically accessible diamond area, i.e. the magnetic field is aligned with the z-axis. At $z = 10$ μm the position of the magnetic field minimum $H(0, 70\ \mu m, 10\ \mu m) = 1100$ A/m is slightly shifted from the loop center due to the effect of the gap. The magnetic field amplitude increases by about a factor of 2 to the edge of optically accessible area. Above the loop the decay of the magnetic field magnitude at the loop center is well described by the decay function for an ideal current loop:

$$H(0,0,z) = H(0,0,0)\frac{D^3}{(4z^2+D^2)^{\frac{3}{2}}} \quad (3)$$

with $D = 440\ \mu m > D_L$ due to the large width of the loop conductor ($W_c = D_L/4$), see Fig. 3(d). Below the loop the magnetic field decays rapidly and is smaller than 70 A/m at the closest position of the microscope objective surface. This guarantees the stability of the system during operation as neither magnetic field distributions nor resonator parameters are influenced by the microscope objective. At higher elevation above the loop the improvement of the magnetic field homogeneity is accompanied by slight loss of the field directivity, see Fig S6 in SI. Within the optically accessible diamond area (±x, ±y, z)=( ±150, ±150, 0+100) µm the minimum and maximum of the magnetic field magnitude are 2200 A/m and 735 A/m at 1 W mw power, respectively. Thus, the magnetic field of the loop compares very favorably with that of a microwire, both in terms of homogeneity and directivity.

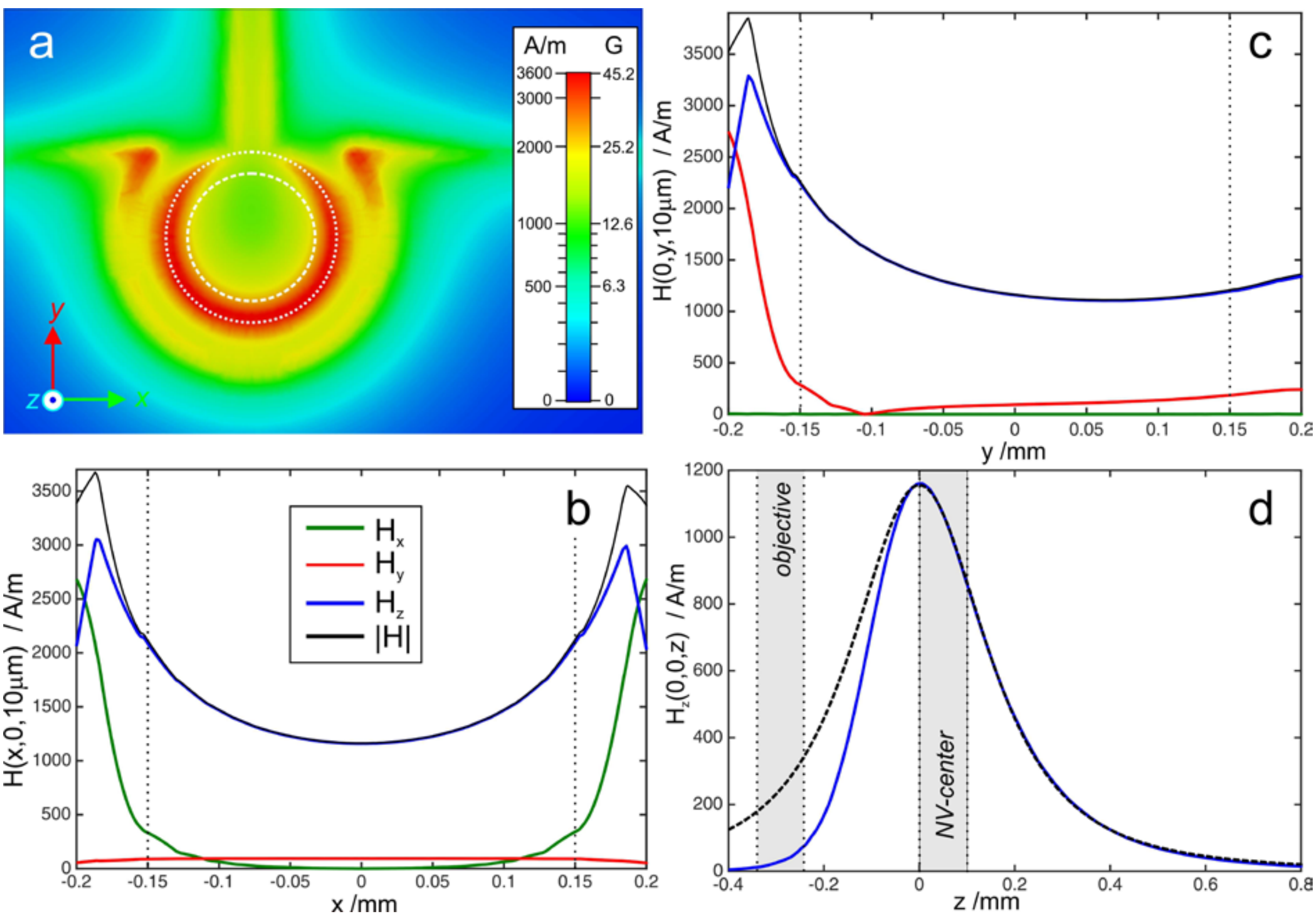

**Figure 3.** **(a)** Color-coded plot of the calculated mw magnetic field magnitude, $|H|$, at $z = 10\ \mu m$ for mw power of 1 W at the resonance frequency. The dotted and dashed circles mark the position of the inner loop edge ($D_L = 400\ \mu m$) and the optical access

hole ($D_{opt} = 300$ µm), respectively. **(b,c,d)** The $H_x, H_y, H_z$ amplitudes of the mw magnetic field components as functions of the position $x, y, z$ for a mw power of 1 W. The black dotted lines mark the optically accessible diamond area. The shadowed area in (d) shows the position region of the optical objective ($z < 0$) and NV-center ($z > 0$). The dashed lines in (d) show the best fit curve of $H_z(0,0,z)$ to the function in Eq. (3).

### 2.3 Reflection mode resonator

The efficiency of the resonator can be improved employing reflection mode. The resonator is converted to reflection mode by just disconnecting the output coax cable. The reflection resonator is described by:

$$R(\nu_0) = \frac{2\sqrt{\beta}}{1+\beta};\ \Delta\nu_{1/2} = \frac{1+\beta}{Q_0} \cdot \nu_0 \tag{4}$$

as derived from the reflection coefficient [16]:

$$\Gamma = -\frac{1-\beta-i\xi}{1+\beta+i\xi}. \tag{5}$$

Thus, for high coupling parameters, $\beta \gg 1$, the reflection mode offers a factor 2 higher magnetic field amplitudes, see SI. The bandwidth, however, becomes reduced by factor of 2. The magnetic field distribution in the loop area is identical in both resonator modes.

Figure 4 shows the calculated and experimental $S_{11}$-parameters of the resonator in reflection mode. Analysis of the S-parameter traces using Eq. (5) yields the resonator parameters $Q_0 = 70$, $\beta = 8.3$ and $Q_0 = 60$, $\beta = 5.8$ for simulation and experiment, respectively. Thus, the bandwidth of the resonator is 400 MHz and 320 MHz for calculation and experiment. The difference between transmission and reflection mode as well as between calculated and experimental reflection parameters are mainly due to the open output line.

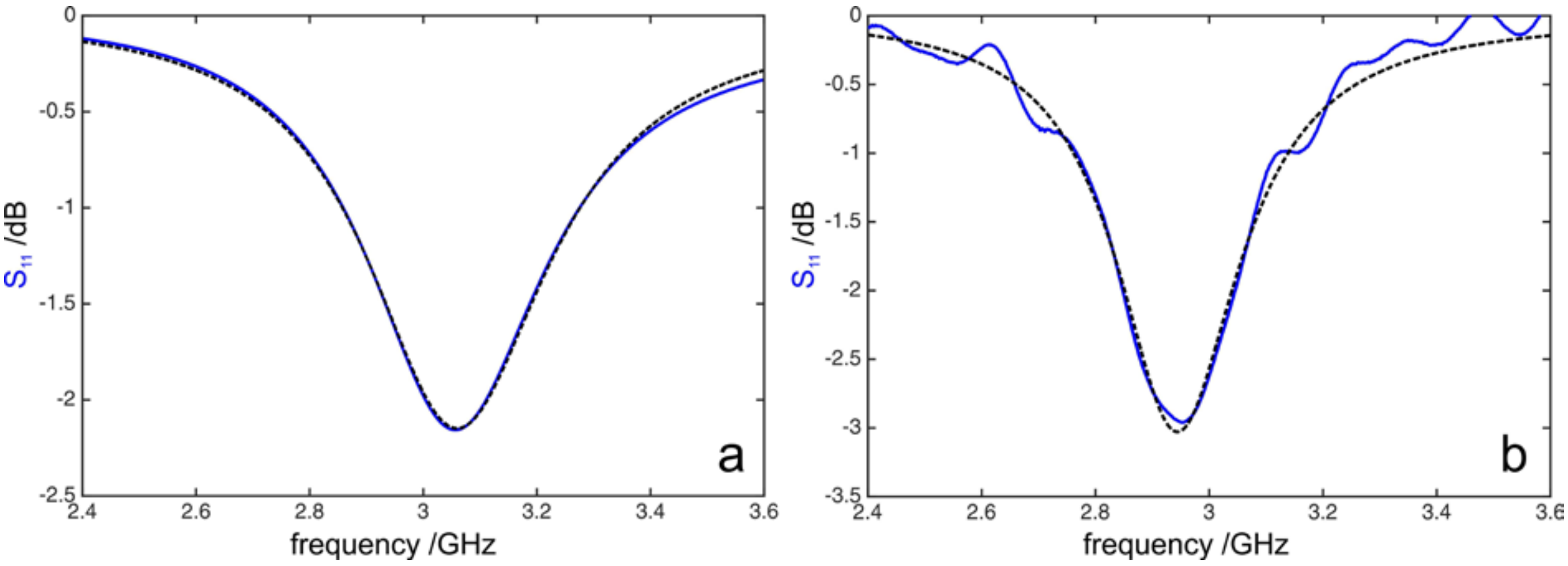

**Figure 4.** Simulated **(a)** and experimental **(b)** $S_{11}$-parameters of the resonator operated in reflection mode. The dashed line show the best fit results of the S-curves to the reflection coefficient given by Eq. (5).

### 3. ODMR experiments

The mw magnetic field behavior was verified experimentally using a previously described ODMR setup capable for continuous wave and pulsed ODMR experiments, see [7] and SI. The $2\times2\times1\,mm^3$ diamond (001 cut) containing a 20 µm layer near surface that was doped with NV centers was fixed to the resonator using transparent office tape, see Figs. S11 and S21 in SI. The central position of the optical objective was set close to the loop center.

Figure 5(a) shows a cw ODMR spectrum of the NV-centers located near the focal spot, recorded using mw power of 6.8 mW (the power level was calibrated at 2.87 GHz). Two intense ODMR lines centered at 2.87 GHz are observed. The line splitting by 475 MHz corresponds to the external magnetic field component of 8.5 mT along the NV-axis. Additional ODMR line pairs with splitting of 260 MHz, 160 MHz and 60 MHz can be attributed to a set of NV-centers within the sensitive volume of about 300 nm in the x-y plane and 1000 nm in the z-direction (see Fig. S9 in SI). The different centers have different orientations along different [1,1,1] directions.

To measure the precise mw magnetic field strength, we recorded Rabi oscillations at the two strongest transitions, as shown in Figure 5(b). The applied mw power of 680 mW resulted in a Rabi frequency of about 14 MHz. The Rabi frequency $f_R$ is proportional to the magnetic field amplitude $H$:

$$f_R = \sqrt{\frac{2}{3}} \cdot \sqrt{2} \cdot \frac{\gamma_e}{2\pi} \frac{\mu_0 H}{2} = \frac{\gamma_e}{2\pi} \frac{\mu_0 H}{\sqrt{3}}, \tag{6}$$

where $\gamma_e$ is the electron gyromagnetic ratio ($\gamma_e/2\pi$=28 GHz/T) and $\mu_0$ is the permeability of the vacuum. The factor $\sqrt{2/3}$ accounts for the component of $H$ that is perpendicular to the NV-axis, (mw H-axis parallel to [0,0,1] and NV-axis parallel to [1,1,1] crystal axis). The factor $\sqrt{2}$ takes into account that we drive one transition of the $S = 1$ spin. Thus, a Rabi frequency of $f_R$=14 MHz at 0.68 W corresponds to a conversion efficiency of $H = 840\,\frac{A}{m}$ at 1 W.

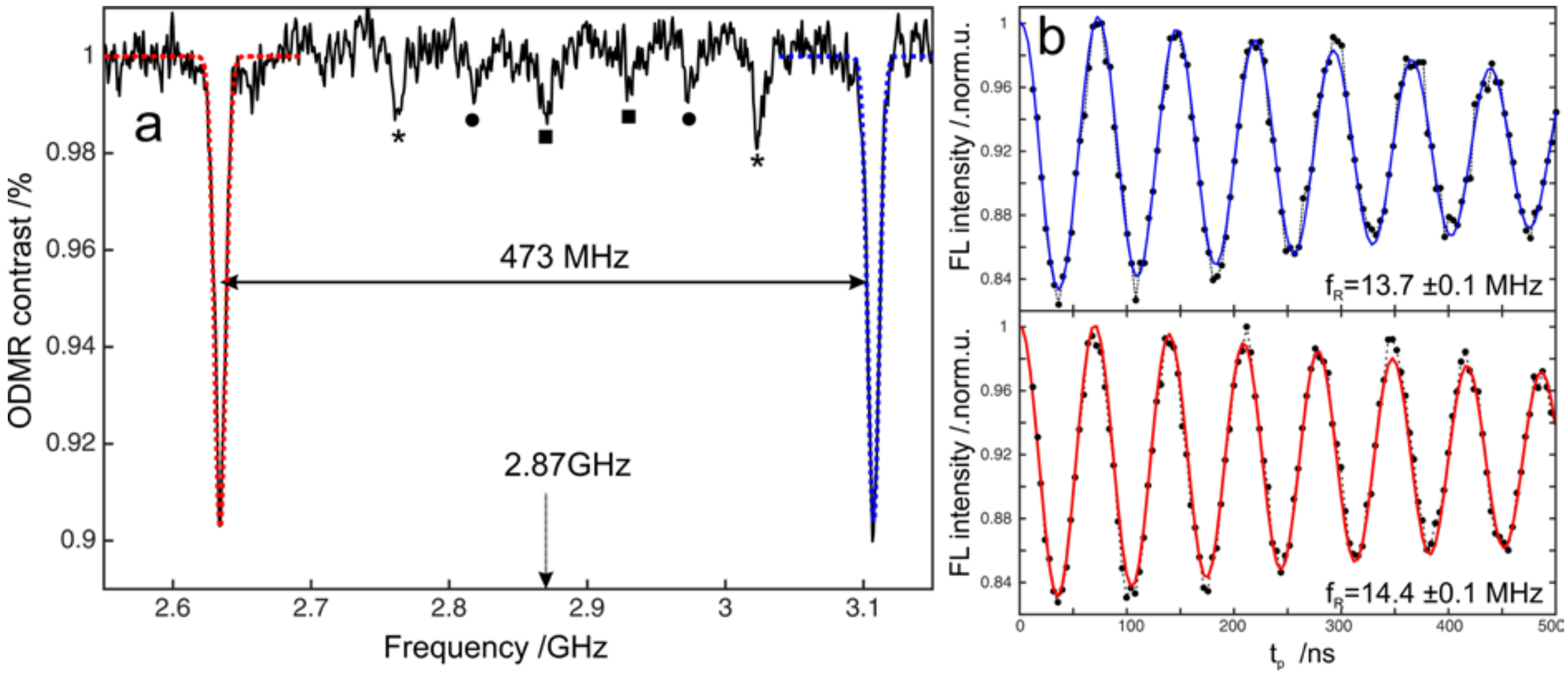

**Figure 5.** **(a)** cw ODMR spectrum of the single NV-center at external magnetic field alight close to [1,1,1] crystal axis recorded at 6.8 mW incident mw power. **(b)** The Rabi oscillation traces recorded at the mw frequencies of the high frequency (upper trace; marked blue in (a)) and low frequency (lower trace; red in (a)) ODMR lines and 680 mW mw power (+20dB increased power over the cw experiment). The solid blue and red lines are the fits.

In the next step the cw and pulsed ODMR experiments were performed at different external magnetic fields and the same input power of the mw amplifier, see Fig. 6(a). The magnetic field amplitude evaluated from the Rabi frequencies overlaid with the simulation results is depicted in Fig. 6(b). Both curves are in the good agreement. The variation of the mw magnetic field over a frequency range of 800 MHz is less than 40%. The discrepancy at the frequencies above 3 GHz is attributed to gain variation of the high-power mw amplifier operated below saturation. The spatial distribution of H in the x-y plane was evaluated from the Rabi experiment at 2.93 GHz for 10 different NV centers over the area of 70×70µm$^2$ accessible by the nanopositioner. The variation of the magnetic field of less than 20% (1000 A/m to 1250 A/m) is in a good agreement with simulation predictions.

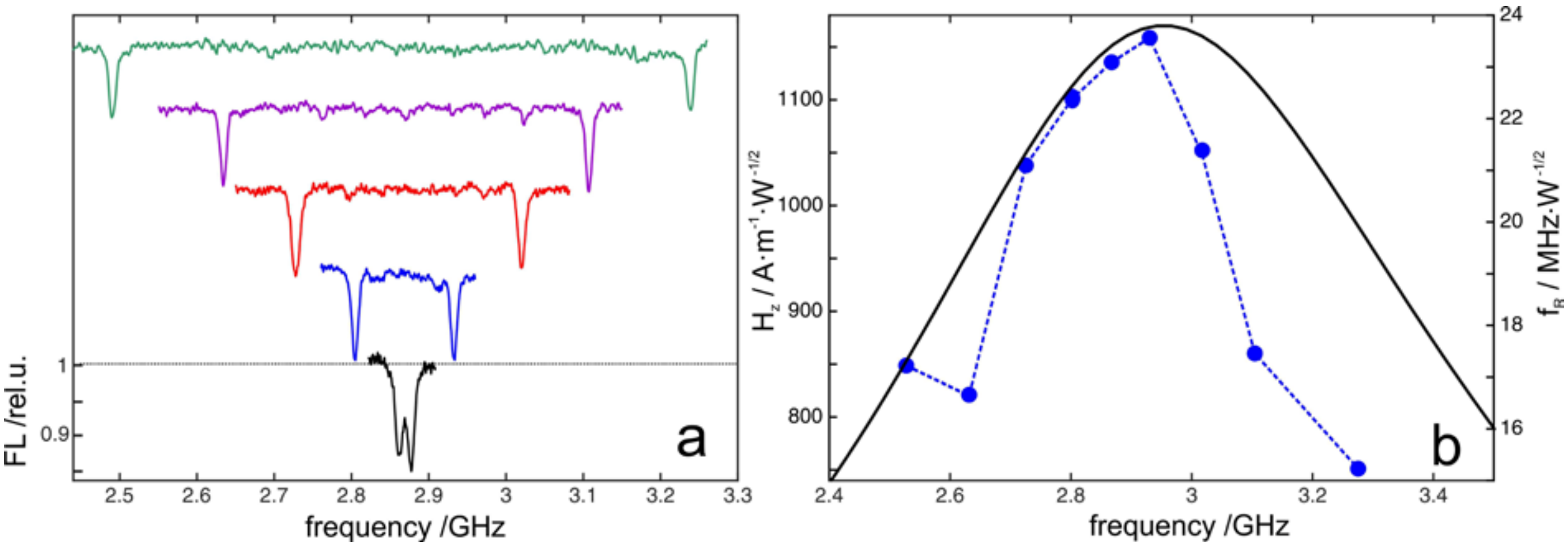

**Figure 6.** **(a)** cw ODMR spectra of the single NV-center at different external magnetic fields at 6.8 mW incident mw power. **(b)** The frequency dependence of the magnetic field

amplitude $H_z(0,0,10\ \mu\text{m})$ calculated for the transmission resonator (black curve, left scale). Rabi nutation frequencies (dots, right scale) evaluated from experimental Rabi oscillation traces recorded at the ODMR line positions in (a). The Rabi frequencies are normalized to 1 W mw power calibrated at 2.87 GHz. The left and right scales are adjusted according to Eq. (6).

These results show that the transmission resonator achieves good spatial magnetic field homogeneity and the bandwidth required for most ODMR experiments on single NV-centers. The transmission resonator achieves π-pulse durations of 20 ns ($f_R = 25$ MHz) using about 1 W mw power. For this performance, a mw amplifier is required. This limitation can be overcome by operating the resonator in reflection mode.

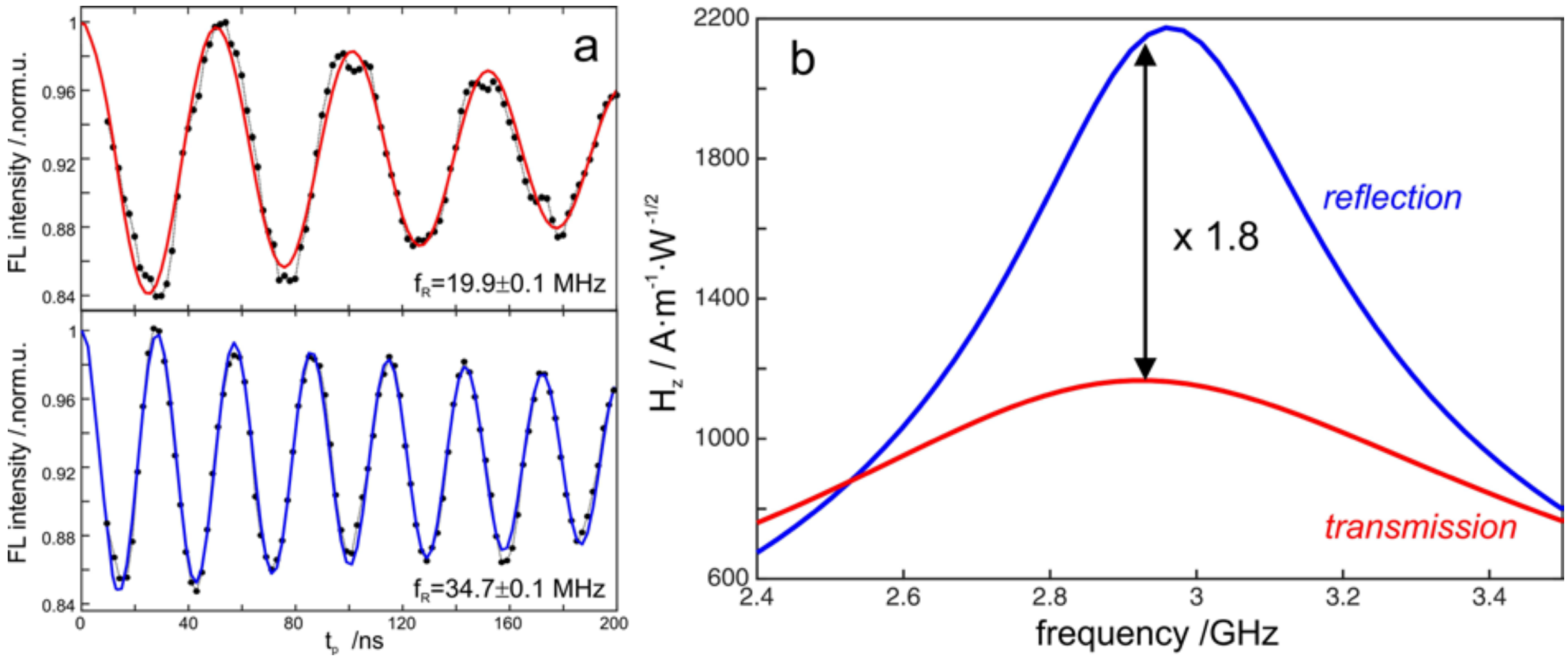

**Figure 7.** **(a)** Rabi oscillation traces recorded at the mw frequency 2.93 GHz for the resonator in transmission (upper trace) and reflection (lower trace) modes at 680 mW mw power. The traces were acquired subsequently at the same settings with the output cable connected and disconnected. The solid-colored lines are the fits. **(b)** Frequency dependence of the magnetic field amplitude $H_z(0,0,10\mu m)$ calculated for transmission (red trace) and reflection resonator (blue trace). The amplitude ratio at 2.93 GHz is indicated by the arrow.

Figure 7(a) shows the Rabi nutation traces recorded at the high frequency ODMR line at 2.93 GHz with 680 mW input mw power. The upper trace was acquired using the resonator in transmission mode. Subsequently, the output mw cable was disconnected converting the resonator into reflection mode and the lower trace was acquired. The magnetic field amplitudes evaluated from Rabi frequencies are 1190 A/m and 2070 A/m for transmission and reflection settings, respectively. The experimentally observed increase of the resonator efficiency by a factor of 1.74 is in a good agreement with factor 1.8 obtained from calculation, see Fig 7(b). Thus, the reflection resonator is capable of producing the same magnetic field amplitude using about factor of 3 smaller mw power. It achieves 20 ns π-pulses ($f_R = 25$ MHz) with as little as 300 mW mw power.

At this point we put two remarks: *(i)* The highest $f_R$ achieved in our setup allows to realize π-pulses with 7 ns duration and 4 ns in transmission and reflection modes, respectively. The pulse profile of such short pulses is not distorted by the resonator, neither in transmission nor in reflection mode, since their voltage ringing times are $t_r = \frac{1}{\pi \Delta \nu_{1/2}} = 0.3$ ns and 1 ns, respectively. *(ii)* The performance of the microscope objective employed in this work, in particular the efficiency of fluorescence collection, is restricted by the resonator acting as an optical diaphragm, see SI. Despite this restriction, the collection of high quality ODMR data is still possible within reasonable time.

**4. Performance summary**

The goal of this work was the design of a device for efficient control of spin quibts. For optimal performance, it should have a number of properties that we specified in the introduction. Here, we first summarise the quantitative performance measures of the mw and rf properties, as shown in Table 1; the values shown were calculated and verified by experiments. For comparison, it also includes the parameters calculated for the 50 Ω microstrip transmission line with the same Ω-type loop. The resonator offers a factor of 4.8 (transmission) and 9.1 (reflection) higher magnetic field amplitudes at the same input mw power as compared to the transmission line. Pulsed ODMR experiments with reasonably short π-pulses of 50 ns can be generated using only 50 mW mw power with the resonator operated in reflection mode. Thus, it allows to avoid high-power mw amplification stages or even to perform the experiments directly using the output of a mw generator. The transmission mode is more advantages for experiments at higher external magnetic fields owing to the larger resonator bandwidth. In contrast to reflection mode, it allows to the perform experiments with combined microwave and radio frequency excitations. An additional advantage of the transmission mode is the good matching of the resonator input. The input return losses are <-20 dB (±50 MHz) and <-10 dB (±150 MHz) around the resonance frequency, see Fig.2. Thus, in contrast to the reflection resonator, the mw excitation system does not require matching elements (circulator or attenuator) on the input for ODMR experiments at small external magnetic fields.

**Table 1.** Summary of resonator and transmission line performances.

| Case | $H_{MW}/\sqrt{P_{in}}$ A/m/$\sqrt{\text{W}}$ | $P_{in}$ [c)] W | $\Delta\nu_{1/2}$ MHz | $H_{RF}/\sqrt{P_{in}}$ [d)] A/m/$\sqrt{\text{W}}$ |
|---|---|---|---|---|
| Transmission line | 245 [a)] | 4.0 | full | 252 (252) |
| Transmission resonator | 1170 [b)] | 0.18 | 950 | 265 (266) |
| Reflection resonator | 2230 [b)] | 0.05 | 400 | 10 (160) |

a) calculated at 2.9 GHz

b) calculated at the resonance frequency (~2.9 GHz)

c) input mw power required for 50 ns π-pulse ($f_R$= 10 MHz) and NV-center in 001 diamond

d) calculated at 10 MHz (200 MHz)

We also considered heating effects. Conduction losses in the loop can lead to a temperature increase that can potentially influence the experiment. The thermal analysis of the transmission mode resonator at ambient temperature shows that the temperature increase in the loop-diamond region is < 0.3 K for 0.1 W continuous mw power (see Fig. S3 in SI). At this mw power the produced magnetic field of 370 A/m corresponds to a Rabi frequency $f_R$=7.5 MHz which is more than sufficient for cw ODMR detection with the highest contrast [6]. In pulsed ODMR experiments the average heat power is significantly smaller. The transmission resonator allows for above $10^6$ π-pulses at 10 W ($t_\pi$= 7 ns) with no significant heating effects. The results of the thermal analysis are consistent with experimental observations.

The resonator design can be adapted for experiments on NV-centers at higher external magnetic fields or different types of optically active spin centers by adjusting some of the design parameters. The variation of the overall resonator length, $L$, between 5 mm and 56 mm allows to tune the resonance frequency between 6 GHz and 1.3 GHz without significant loss of the mw performance, see Fig. S16 in SI. Further increasing or decreasing the resonance frequency is possible by adjusting the dielectric constant and thickness of the substrate, and the geometry of the holder.

Various types of microwave resonators have been proposed for experiments on optically active spin centers, such as the diamond NV center, each with specific advantages and limitations. A fair comparison of the performance of all these different designs would have to be made under a specific set of conditions, where all designs can operate. Since this is not possible, we include here a comparison with two recently reported designs with overlapping boundary conditions, which are applied in several laboratories for experiments on NV-centers [10, 11, 13, 19-26]. The first system, originally reported by Bayat et al. [10], is based on a double-split-ring resonator operated in reflection. In original design has an efficiency of 355 A/m/$\sqrt{\mathrm{W}}$ , with a bandwidth of 40 MHz. Similar numbers were reported for some modified designs [11, 25]. The second design, originally reported by Sasaki et al. [13], is based on a split-ring (or loop-gap) resonator in reflection mode. In original design provides 240 A/m/$\sqrt{\mathrm{W}}$ with a bandwidth of 440 MHz. Better parameters 485 A/m/$\sqrt{\mathrm{W}}$ and 300 MHz are reported for the modified version [20]. Thus, the performance of our system in both transmission and reflection modes of operation is superior to both systems.

## 5. Conclusion

In this work we presented a novel mw excitation system based on a resonator designed for cw and pulsed ODMR experiments on single NV-centers to be combined with confocal microscopy. The high performance of the system was verified using numerical EM calculations and ODMR experiments. The resonator can be easily and cheaply fabricated using standard PCB lithography. It offers high mechanical and thermal stability. Its performance in terms of mw power to magnetic field conversion efficiency and magnetic field homogeneity is superior in comparison to currently used systems like the microwire-transmission line system. It also offers a range of practical advantages such as simplified optical adjustment and optimization of the mw power settings for all types of ODMR experiments. It also offers the possibility for quick replacement of the diamond crystal because it does not have to be permanently connected to the resonator. This provides the opportunity for investigation of different samples with the same structure or use a resonator which is optimized for the performance of a specific type of ODMR experiment.

While the present work focuses on applications on NV centers in diamond, the principles used here are completely general and can be readily transferred to similar systems that rely on efficient spin control by microwave fields. This includes not only ensembles of NV centers but also other materials like semiconductors.

## Acknowledgments

This project has received funding from the European Union's Horizon 2020 research and innovation program under grant agreement No 828946. The publication reflects the opinion of the authors; the agency and the commission may not be held responsible for the information contained in it.

## Author Declarations

### Conflict of Interest

The authors have no conflicts to disclose.

### Data Availability

The data that support the findings of this study are available from the corresponding authors upon reasonable request.

### Author Contributions

**Anton Savitsky:** Conceptualization (equal); Formal analysis, (lead); Investigation (equal); Methodology (lead); Visualization (lead); Writing – original draft (lead). **Jingfu Zhang:** Data curation (equal); Investigation (equal); Formal analysis (equal); **Dieter Suter:** Conceptualization (equal); Review & editing (equal); Funding acquisition (lead).

# Supporting Information

## Variable bandwidth, high efficiency microwave resonator for control of spin-qubits in nitrogen-vacancy centers

Anton Savitsky, Jingfu Zhang and Dieter Suter

*Faculty of Physics, Technical University Dortmund, Otto-Hahn-Str. 4a, 44227 Dortmund, Germany*

**Content:**

## S1 Optimization of the resonator parameters

In this part we consider some important points for the resonator optimization based on the first principals and simulations. In the first part we consider transmission line with Ω-type loop which allows to optimize parameters of the loop. In the second part we compare the performance of resonator with transmission and reflection line and point out the ways for performance optimization.

### S1.1 Optimization of Ω-type loop parameters

The loop of diameter $D$ integrated in the gap of a terminated microstrip transmission line can be considered as an approximation of the ideal classical current loop. This approximation is valid when *(i)* the loop length is much smaller than a quarter of the wavelength ($\pi D \ll \lambda_g/4$); *(ii)* the width of the loop is much smaller compared to it diameter, *(iii)* the shielding effects can be neglected; and *(iv)* the loop-gap discontinuity is neglected. The EM-wave of power $P_{in}$ in the microstrip line with $Z_L$ generates an alternating electric current with the amplitude

$$I_{loop} = \sqrt{\frac{2P_{in}}{Z_L}}.$$

This alternating current induces a magnetic field whose amplitude at the center of the loop is

$$H(0,0,0) = \frac{I_{loop}}{D} = \frac{1}{D}\sqrt{\frac{2P_{in}}{Z_L}}.$$

The loop conversion efficiency

$$H_0 = \frac{H(0,0,0)}{\sqrt{P_{in}}} = \frac{1}{D}\sqrt{\frac{2}{Z_L}}$$

can be increased either by decreasing the loop diameter or by lowering the line impedance. The line impedance of $Z_L = 50\ \Omega$ is, however, generally fixed to avoid mismatch with microwave components. This results in $H_0 = 500\ \frac{A}{m\sqrt{W}}$ for the loop of $D = 400$ µm used here. This value defines the physical upper limit of the magnetic field amplitude reachable for a transmission line.

The real Ω-loop differs from the ideal loop. There are several parameters which influence the magnetic field amplitude and its distribution: *(i)* loop width, *(ii)* gap width and *(iii)* thickness of the dielectric substrate. The last factor in our system is, however, defined by the necessity to provide the optical access to the diamond surface and will be not considered here, i.e. we fix the substrate and optical access parameters defined in the main text, see Fig. 1 (main text), and consider only the first two parameters.

***Loop conductor width.*** The width of the loop is of critical importance as it influences not only magnetic field parameters but also the heating effects. Figure S1(a) shows the dependence of the conversion efficiency on the loop width. As expected, the conversion efficiency decreases with increasing loop width. The maximum efficiency is about a factor 1.6 smaller than in the ideal loop due to the effect of the gap and metal shield on the back side of the structure. The conductive power losses, $P_{loss} = \frac{I_{loop}^2}{2} \cdot R_{loop}$, in the loop also decrease, see Fig S1(b), due to the decrease of the loop resistance, $R_{loop}$. The optimal loop width of $W_C = 0.1$ mm is estimated by considering the maximum of the function $H_0^2/P_{loop}$ , see Fig S2. The conversion efficiency drops to 245 $\frac{A}{m\sqrt{W}}$, i.e. a factor 2 smaller than for the ideal loop. The loop width of 0.1 mm also guarantees better mechanical stability and stability of the system parameters to manufacturing tolerances.

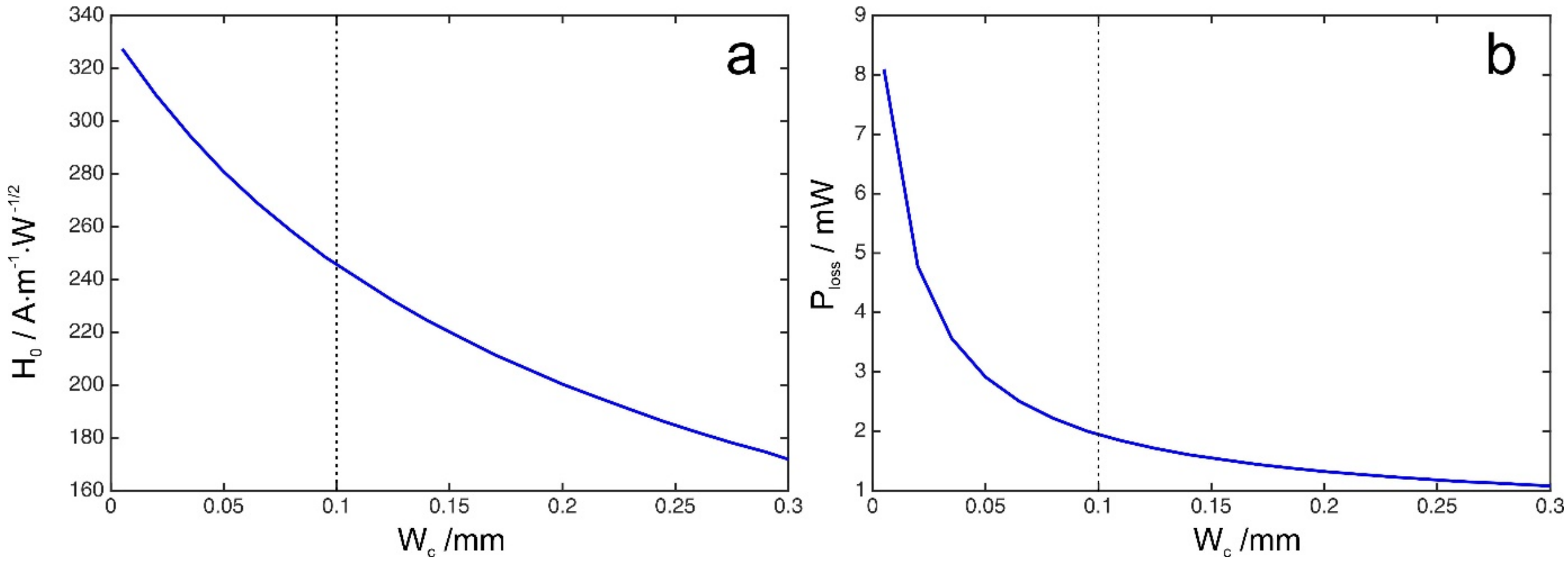

**Figure S1.** **(a)** Conversion efficiency and **(b)** loop ohmic losses calculated for different loop widths and $Z_L = 50\ \Omega$ transmission line feeded with 1 W mw power at 3 GHz. The gap width is G =0.1 mm. The dashed line marks the value $W_C$ chosen for fabrication.

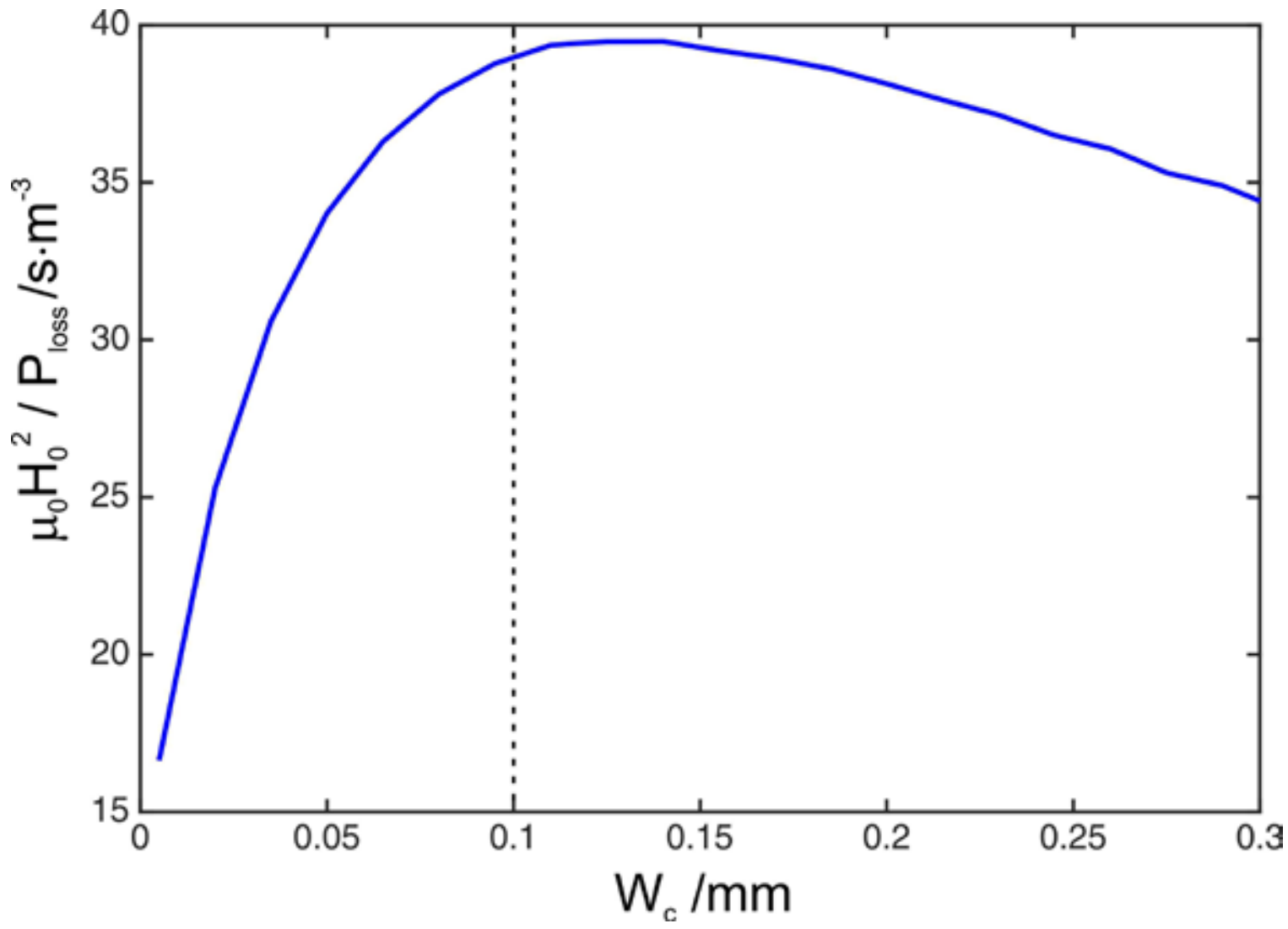

**Figure S2.** Ratio of power conversion efficiency to conductive losses in the loop.

In order to estimate the heating effects we have calculated the temperature distribution in the transmission resonator continuously fed by mw power, see Fig. S3. For $W_C = 0.1$ mm the maximum temperature calculated in the system is about 0.3 K (0.1 W) and 3.5 K (1 W) starting from the ambient temperature of 293 K. In contrast, the system with $W_C = 0.018$ mm shows higher temperature increase of 0.6 K (0.1 W) and 5.8 K (1 W excitation) due to the higher ohmic resistance and lower thermal conductance of the loop.

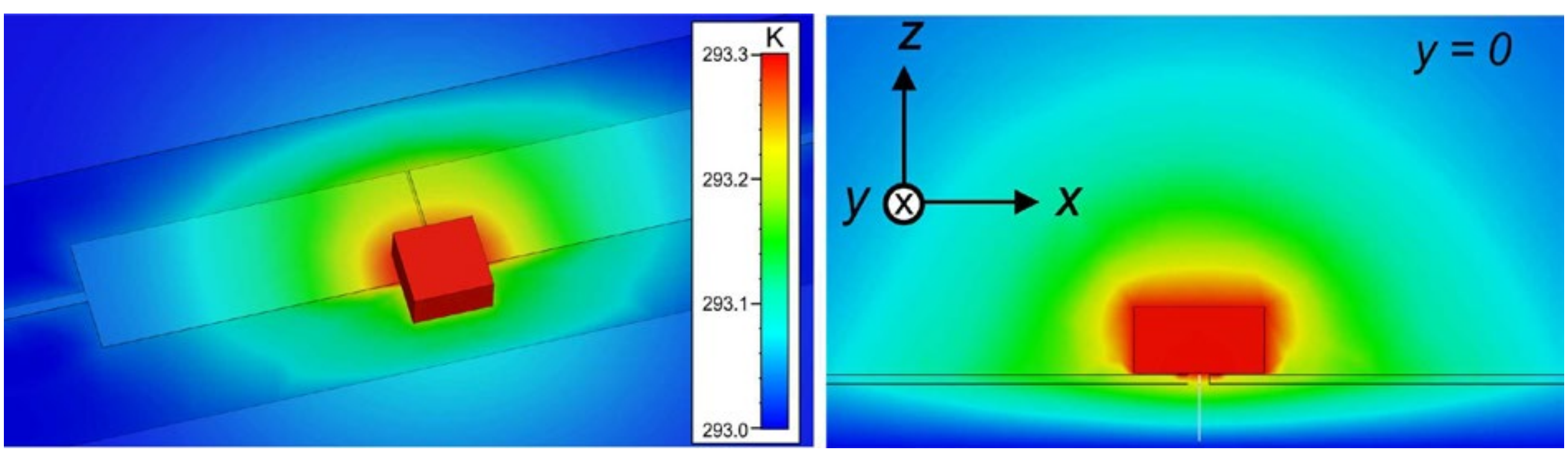

**Figure S3.** Temperature distribution calculated for the transmission resonator including the diamond crystal (2×2×1 mm$^3$) with geometrical parameters given in Fig. 1 (main text). CW input mw power was set to 0.1 W.

***Gap width.*** Figure S4 shows the efficiency parameter as a function of the gap width. Increasing the gap width from 0.005 mm to 0.3 mm ($G = 0.75 \cdot D_L$) has a small effect on the resonator efficiency (magnetic field amplitude at the loop center) as well as the field distribution in y < 0 loop region. Therefore, for our resonator we have chosen G = 0.1 mm for manufacturing ease.

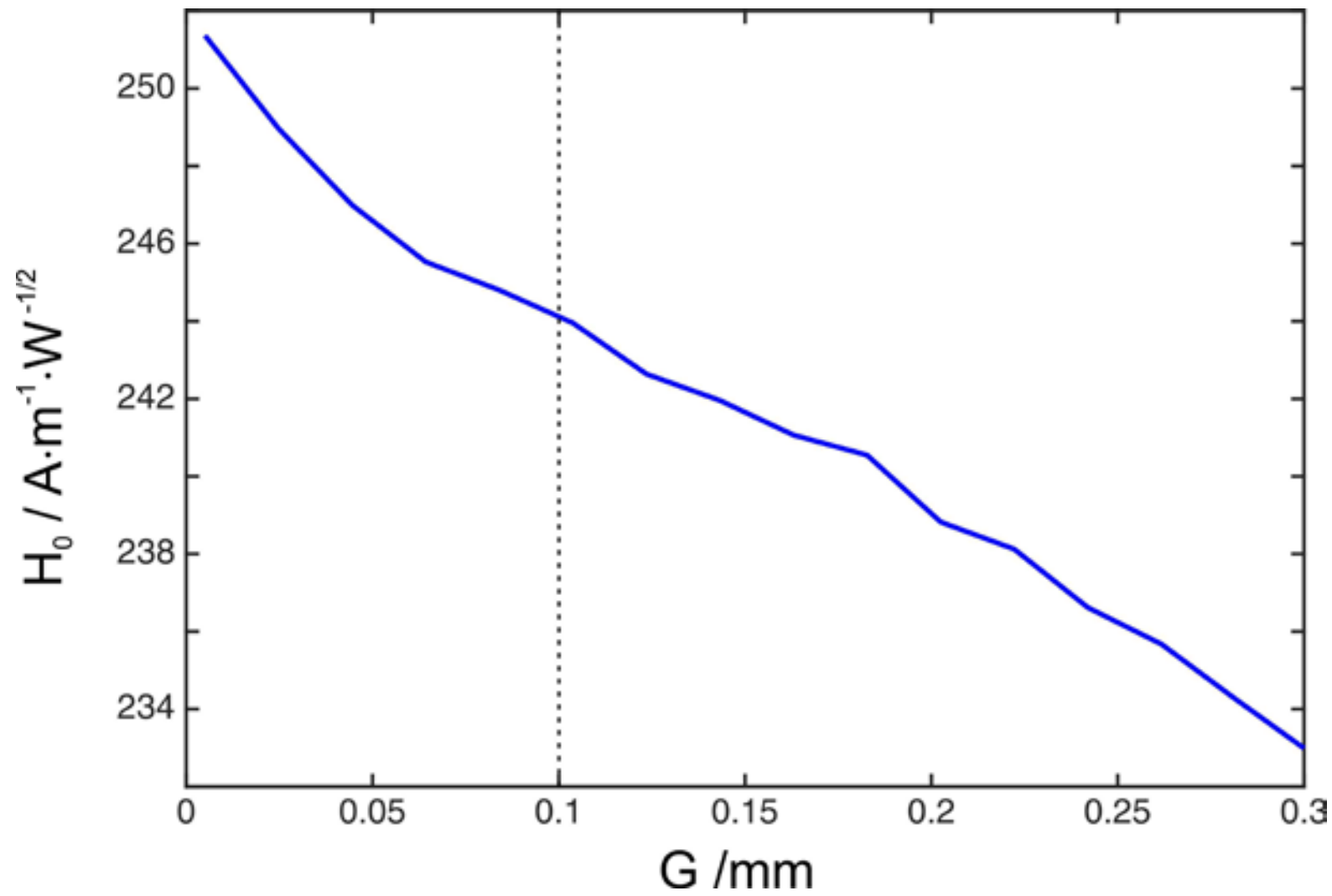

**Figure S4.** Conversion efficiency calculated for different gap widths $G$ and $Z_L = 50\Omega$ transmission line fed with 1 W mw power at 3 GHz. The loop width is $W_L$= 0.1 mm.

### S1.2 Calculation of the coupling parameter for width step coupling

The relation between the properties of the coupling element and the resonator coupling parameter can be derived in two following steps.

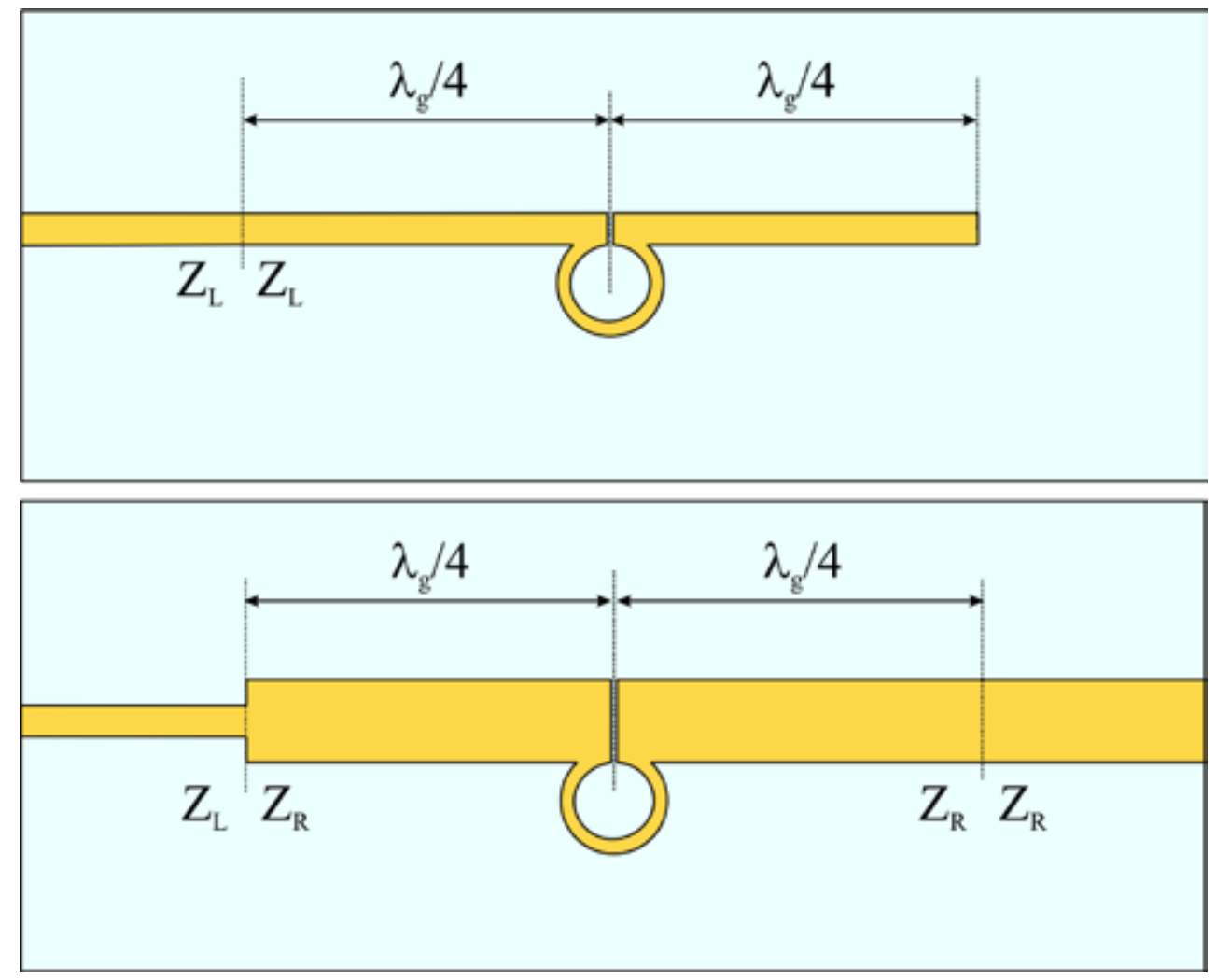

**Figure S5. (a)** Reflection line with Ω-type loop. **(b)** Transmission line with width step and Ω-type loop.

***Reflection line.*** The bandwidth of reflection line, Fig S5(a), can be determined considering the frequency dependence of magnetic field in the loop center:

$$\frac{H(\nu)}{H(\nu_0 = \frac{\lambda_g}{c})} = \sin\left(2\pi\nu\frac{\lambda_g}{4c}\right) = \frac{1}{\sqrt{2}}$$

The magnetic field amplitude is reduced by $\sqrt{2}$ at $\nu = \nu_0/2$ which results in the bandwidth

$$\Delta\nu_{1/2} = \nu_0$$

The reflection line can be also considered as overcoupled resonator with the bandwidth of

$$\Delta\nu_{1/2} = \frac{1+\beta}{Q_0} \cdot \nu_0$$

Thus, the coupling coefficient of the reflection resonator with $Z_R = Z_L$ is given by

$$\beta = Q_0 - 1 \approx Q_0\ (for\ Q_0 \gg 1)$$

The same result can be also derived from the magnetic field amplitude than considering reflection line as the reflection resonator

$$\frac{2\sqrt{\beta}}{1+\beta}\sqrt{Q_0} = 2 \Rightarrow \sqrt{\beta} = \frac{\sqrt{Q_0} + \sqrt{Q_0 - 4}}{2} \Rightarrow \beta \approx Q_0$$

Generally the result is quite obvious than consider the definition of the coupling coefficient

$$\beta = \frac{Q_0}{Q_E}$$

Thus, the external quality factor in reflection line $Q_E = 1$ or full power is dissipated in external load than neglecting resonator losses, $Q_0 \gg 1$.

***Transmission line with width step.*** This system can be considered as transmission resonator, see Fig. S5(b). The reflection coefficient at $\nu = \nu_0$ of this system with $\beta' = Q_0 - 1$ is given by

$$\Gamma = -\frac{1-\beta+\beta'}{1+\beta+\beta'} = -\frac{-\beta+Q_0}{\beta+Q_0} \Rightarrow \beta = Q_0\frac{\Gamma-1}{\Gamma+1}$$

Taking into account that, the reflection coefficients of width step discontinuity is

$$\Gamma = -\frac{Z_L - Z_R}{Z_L + Z_R}$$

one obtains

$$\beta = Q_0\frac{Z_R}{Z_L} \text{ or } Q_E = \frac{Z_L}{Z_R}$$

Thus, the bandwidth of the resonator can be calculated

$$reflection\text{:}\ \frac{\Delta\nu_{1/2}}{\nu_0} = \frac{1+\beta}{Q_0} = \frac{1}{Q_0} + \frac{Z_R}{Z_L}$$

$$transmission\text{:}\ \frac{\Delta\nu_{1/2}}{\nu_0} = \frac{1+2\beta}{Q_0} = \frac{1}{Q_0} + 2\frac{Z_R}{Z_L}$$

For the resonator design in the main text ($Z_L = 50\ \Omega;\ \ Q_0 = 74$) the above equation yields the estimate value of the coupling parameter of $\beta = 12.6$, than assuming the resonator impedance to be equal to that of the transmission microstrip line with width $W_R$. The values is in a good agreement with coupling parameter of 11.5 evaluated from simulated S-parameters.

**S1.3 Performance comparison of the resonator with transmission, reflection line**

The conversion efficiency for transmission, reflection resonator with ideal loop can be calculated using following relations

$$reflection\text{:}\ H_0 = \frac{2\sqrt{\beta}}{1+\beta}\cdot\sqrt{Q_0}\cdot\frac{1}{D}\sqrt{\frac{2}{Z_R}}$$

$$transmission\text{:}\ H_0 = \frac{2\sqrt{\beta}}{1+2\beta}\cdot\sqrt{Q_0}\cdot\frac{1}{D}\sqrt{\frac{2}{Z_R}}$$

They take into account *(i)* power flow to resonator (coupling); *(ii)* the energy storage in the resonator (quality factor); and *(iii)* the increase of the loop current due to decrease of the resonator impedance. It is important that for high coupling coefficients ($\beta = Q_0\frac{Z_R}{Z_L} \gg 1$) both factors become independent of resonator quality factor, see Table S1. Under this condition the increase of the resonator efficiency as compared to the transmission line can be estimated from the ratio of the resonator and feed line impedances, i.e. $\frac{Z_L}{Z_R}$ for transmission and $2 \cdot \frac{Z_L}{Z_R}$ for reflection. For instance, the resonator having $Z_R = 10\ \Omega$ is factor 5 more efficient compared to transmission line having the same loop, i.e. it requires 25 times less mw power to obtain the same magnetic field amplitude. We note that $Z_R$ can be approximated by the impedance of the microstrip line with width $W_R$ only than the discontinuity contribution to impedance can be neglected. This generally holds for $W_R \ll L$. In general case $Z_R$ have to include the discontinuity contribution which is best evaluated from numerical simulation. It is important to note that then the heating effects become the limiting factor, the excitation mw power have to be limited due to $P_{loss} \propto I_{loop}^2$ and, as the result, the same maximum magnetic field is reachable in all cases.

**Table S1.** Analytical expressions for efficiency and the bandwidth of the resonator and non-resonating systems containing ideal current loop.

| Case | $H_0$ | $\frac{\Delta\nu_{1/2}}{\nu_0}$ | $H_0$ | $\frac{\Delta\nu_{1/2}}{\nu_0}$ |
|---|---|---|---|---|
| Transmission line | $\frac{1}{D}\sqrt{\frac{2}{Z_L}}$ | $full$ | $\frac{1}{D}\sqrt{\frac{2}{Z_L}}$ | $full$ |
| Reflection line | $2 \cdot \frac{1}{D}\sqrt{\frac{2}{Z_L}}$ | 1 | $2 \cdot \frac{1}{D}\sqrt{\frac{2}{Z_L}}$ | 1 |
| Transmission resonator | $\frac{2\sqrt{\beta}}{1+2\beta}\sqrt{Q_0}\frac{1}{D}\sqrt{\frac{2}{Z_R}}$ | $\frac{1+2\beta}{Q_0}$ | $\frac{Z_L}{Z_R} \cdot \frac{1}{D}\sqrt{\frac{2}{Z_L}}$ | $2\frac{Z_R}{Z_L}$ |
| Reflection resonator | $\frac{2\sqrt{\beta}}{1+\beta}\sqrt{Q_0}\frac{1}{D}\sqrt{\frac{2}{Z_R}}$ | $\frac{1+\beta}{Q_0}$ | $\frac{2Z_L}{Z_R} \cdot \frac{1}{D}\sqrt{\frac{2}{Z_L}}$ | $\frac{Z_R}{Z_L}$ |

Table S2 summarizes the parameters of the transmission resonator calculated for different $W_R$ and $L$ values. At large $W_R$'s the contribution of the loop and gap to resonator impedance overcomes the impedance contribution of the microstrip line. Thus, it becomes the limiting factor determining resonator performance, i.e. conversion factor and the bandwidth.

**Table S2.** Parameters of transmission resonator calculated for different $W_R$ and $L$, for definition see Fig. 1 in main text.

| Case [a] | $H_0$ A/m/√W | $H_0^R/H_0^{TL}$ [b] | $\beta$ [c] | $Q_0$ [c] | $\frac{\Delta\nu_{1/2}}{\nu_0}$ | $Z_R$ [d] Ω | $Z$ [e] Ω |
|---|---|---|---|---|---|---|---|
| $W_R = 1\ mm$ $L = 24.5\ mm$ | 576 | 2.35 | 32.9 | 92 | 0.73 | 21.3 | 21.6 |
| $W_R = 2\ mm$ $L = 22\ mm$ | 930 | 3.79 | 14.67 | 79 | 0.38 | 13.2 | 12.1 |
| $W_R = 3\ mm$ $L = 17\ mm$ | 1170 | 4.80 | 11.50 | 74 | 0.32 | 10.4 | 8.4 |
| $W_R = 4\ mm$ $L = 15\ mm$ | 1300 | 5.33 | 8.77 | 72 | 0.26 | 9.4 | 6.4 |
| $W_R = 5\ mm$ $L = 13\ mm$ | 1379 | 5.65 | 7.90 | 69 | 0.24 | 8.8 | 5.2 |
| $W_R = 5.6\ mm$ $L = 11.2\ mm$ | 1386 | 5.68 | 7.65 | 72 | 0.23 | 8.8 | 4.7 |
| $W_R = 6\ mm$ $L = 10\ mm$ | 1422 | 5.82 | 7.54 | 62 | 0.26 | 8.6 | 4.4 |

a) $L$ was adjusted to obtain the resonance frequency of 2.95±0.05 GHz
b) Resonator efficiency relative to the efficiency of the transmission line (245 A/m/√W)
c) Evaluated from $S_{11}$ and $S_{21}$ curves
d) Resonator impedance calculated as $Z_R = Z_L/(H_0^R/H_0^{TL})$
e) Impedance of microstrip line with width $W_R$.

### S2. MW magnetic field distribution at z = 100 μm

Figure S6 shows the mw magnetic field distribution at $z$ = 100 μm above the loop, i.e. about the highest optically accessible position with our optical objective with 340 μm working distance. Compared to $z$ = 10 μm (Fig. 3 main text) the field homogeneity is improved. Over the full optically accessible area the maximum and minimum field magnitudes of 1020 $A/m$ and 735 $A/m$ are calculated, i.e. 735 $A/m < |H(x,y,100\mu m)| < 1020\ A/m$ (for comparison 1070 $A/m < |H(x,y,10\mu m)| < 2270 A/m$). For the loop center and area accessible by the nanopositioner (70×70μm$^2$) the field homogeneity is even better, i.e. 832 $A/m < |H(x,y,100\mu m)| < 908\ A/m$ ($1085\ A/m < |H(x,y,10\mu m)| < 1256\ A/m$). The magnetic field directivity is not as perfect as at small elevations but still sufficiently high to be neglected, especially in the loop center. The fields distribution compares favorably to that reported previously for planar loop-gap resonator [1].

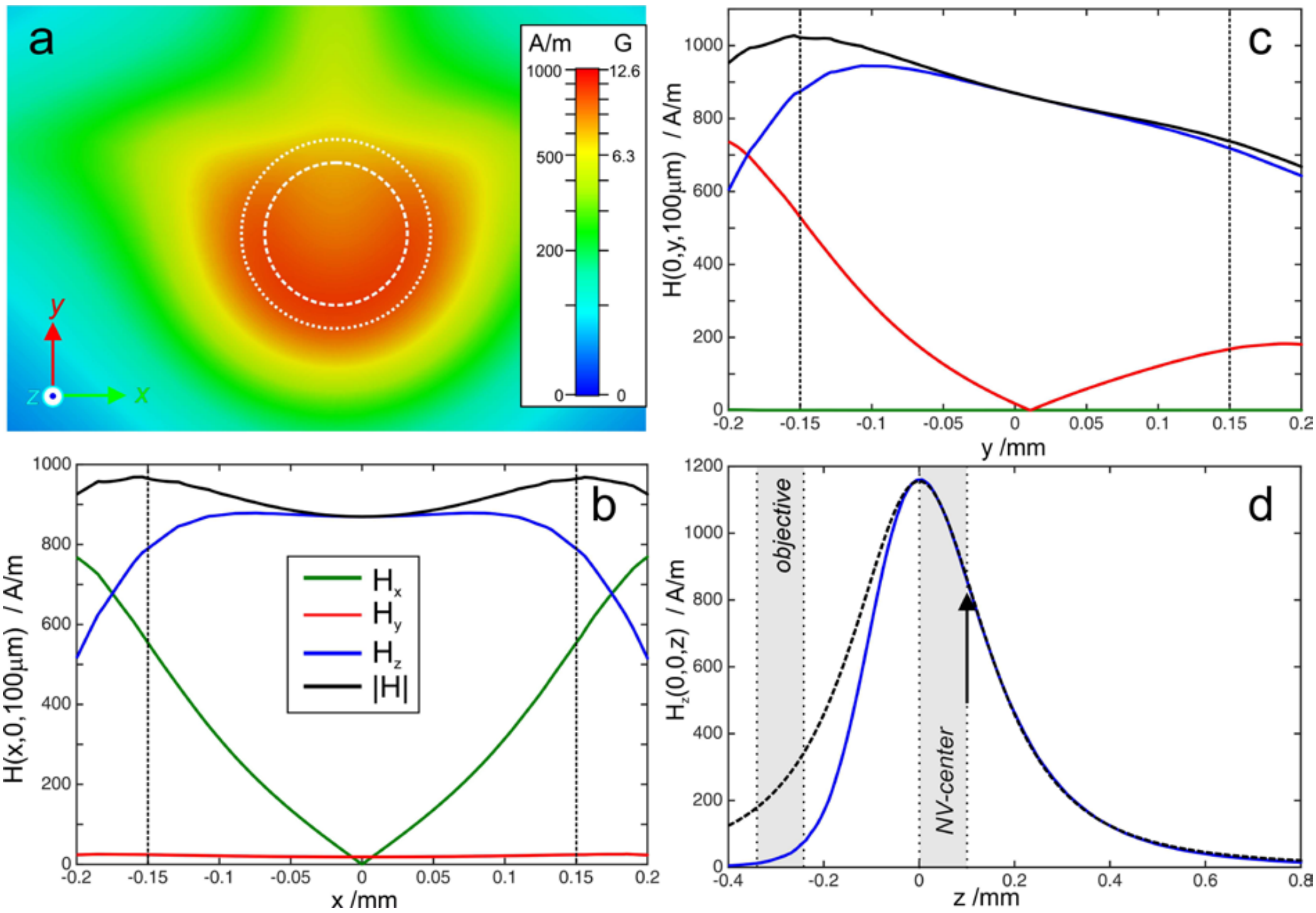

**Figure S6. (a)** Calculated microwave magnetic field magnitude, |H|, at $z$ = 100 μm for a mw power of 1 W. The dotted and dashed circles mark the position of the inner loop edge ($D_L = 400 \mu m$) and the optical access hole ($D_{opt}$ = 300 μm), respectively. **(b,c,d)** The $H_x, H_y, H_z$ amplitudes of the mw magnetic field components as functions of x,y,z for mw power of 1 W. The black dotted lines mark the optically accessible diamond area. Shadowed area in (d) shows the position region of the optical objective ($z < 0$) and NV-center ($z > 0$). The dashed curve in (d) shows the best fit curve of $H_z(0,0,z)$ amplitude to the function in Eq. (3) in the main text.

## S3. Radio frequency performance

In experiments where Radio frequency (RF) is used to drive nuclear spins, the corresponding signals can be fed into the same resonator. Figure S7 shows the power-to-field conversion efficiency of the different types for RF signals for the frequency range < 200 MHz. The structures (transmission line, mw resonator) are non-resonant at RF fields and therefore relatively broad-band. In transmission, the efficiency does not depend on the frequency. The reflection resonator performs poorly at low frequencies and should not be used under such conditions.

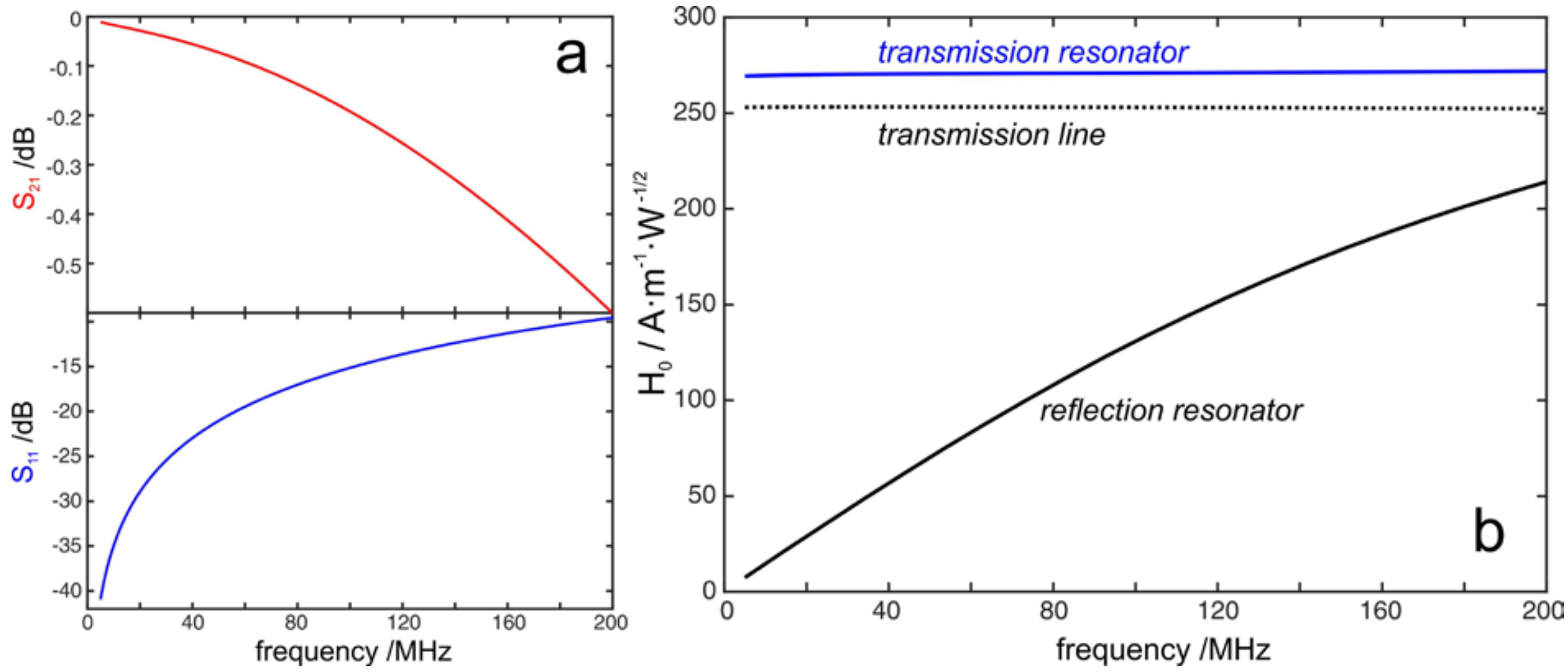

**Figure S7.** **(a)** Simulated $S_{21}$ (upper trace) and $S_{11}$ (lower trace) of the transmission resonator with diamond for the radio frequency range. **(b)** The frequency dependence of the magnetic field amplitude $H_z(0,0,10\ \mu\text{m})$ calculated for the transmission line and the resonator in reflection and transmission modes.

## S4. Fluorescence collection efficiency and optical resolution

The fluorescence collection efficiency for an NV center as the point emitter and neglecting reflections on diamond-oil interface is

$$FCE = \frac{1 - \cos\alpha}{2},$$

The light cone opening angle, $\alpha$, of the optical objective employed in this work (Zeiss, Apochromat 100 with $NA = 1.4$; immersing oil Immersol 518 F with $n = 1.518$) is $\alpha = \text{asin}(\frac{NA}{n}) = 67°$. Thus, the optically unrestricted objective is capable of collecting 30% of the emitted photons. The diameter of the optical access hole required for this efficiency would be $D_{opt} = 0.78\ mm$ (loop inner diameter $D_{in} = 0.88\ mm$) for NV centers close to the diamond surface and a resonator thickness (substrate and copper cladding) of 0.166 mm, see Fig. S8(c). In this work we decided for $D_{opt} = 0.3\ mm$ allowing for $\alpha = 42°$, see Fig. S8(a), in order to optimize the microwave performance of the resonator. The collection efficiency, however, becomes potentially reduced by factor of 2.4 as compared to the maximum possible for this objective. Increasing the optical access hole diameter will improve the collection efficiency, see Fig. S8(b) at the cost of *(i)* reduced power to magnetic field conversion efficiency, see Table S1; *(ii)* increased heating; *(iii)* increased effect of the objective on the properties of the resonator (resonance frequency and magnetic field distribution). These factors have to be taken into account when optimizing the resonator geometry for experiments which require high collection efficiency.

α

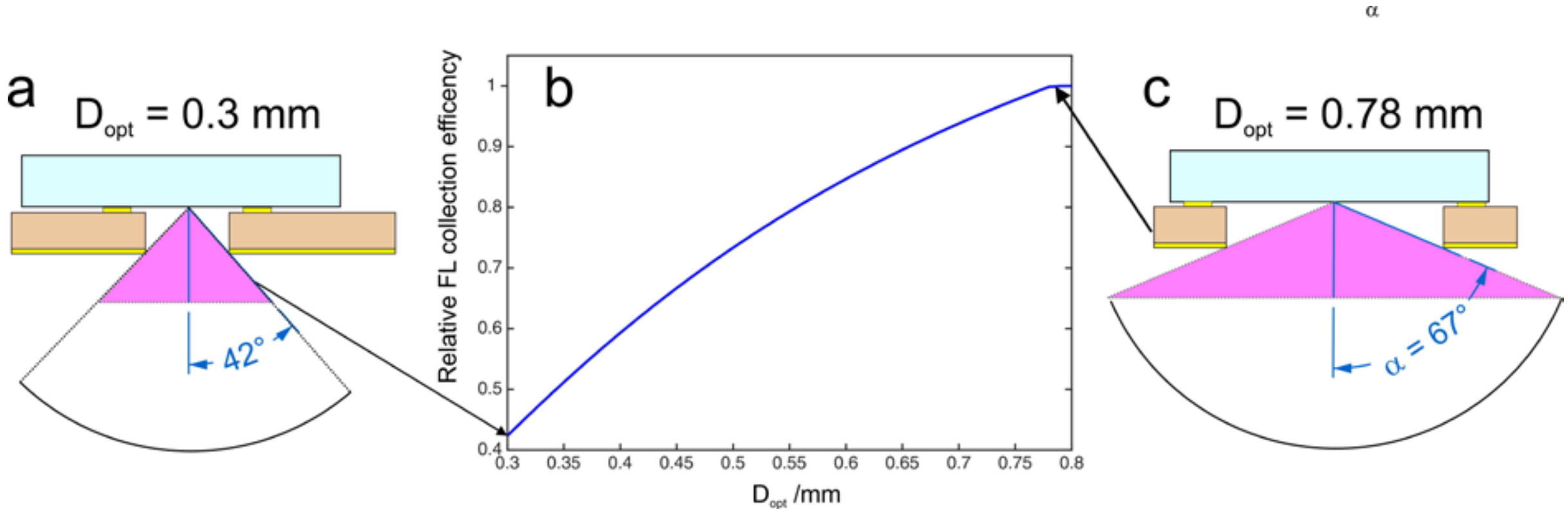

**Figure S8.** Fluorescence optical pathway for **(a)** the resonator employed in this work and **(c)** resonator configuration required for maximum fluorescence collection efficiency. **(b)** Dependence of fluorescence collection efficiency on diameter of optical access hole relative to that capable by objective.

***Optical resolution.*** As pointed above the resonator acts as a diaphragm reducing numerical apperture of the objective and, as the result, limiting the optical resolution. The effect is, however, weaker compared to the loss of the fluorecence collection efficiency. Figure S9(a) shows the fluorecence image of the single NV-center recorded with the resonator setup. Evaluation of the fluorecence spatial distribution yields the FWHM of recordings about 300 nm, Fig. S9(b). This number corresponds to $300nm\frac{\sin(42°)}{\sin(67°)} = 220\ nm$ for objective without diaphragm, which is close to the number calculated for Airy pattern of a point emmiter $\frac{0.51\lambda}{NA} = \frac{0.51\cdot 700nm}{1.4} = 255\ nm.$

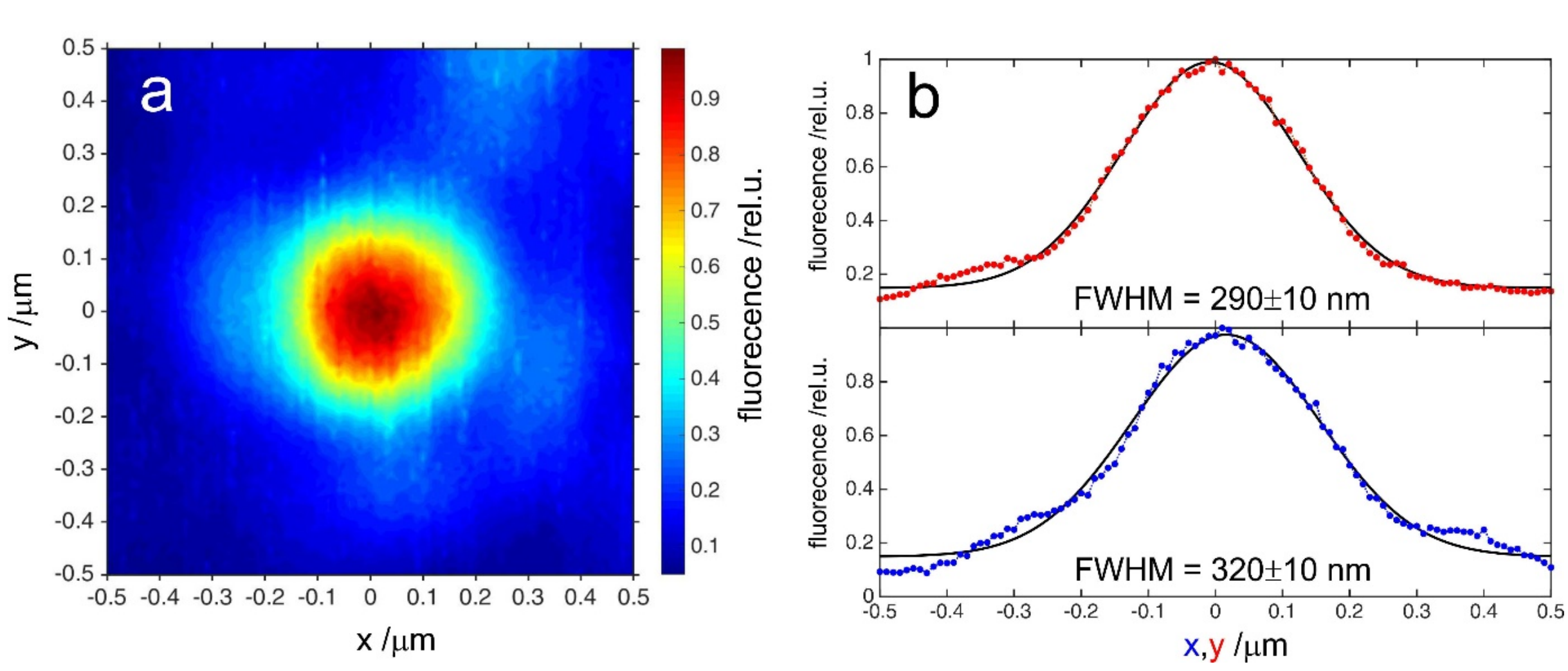

**Figure S9.** Optical resolution. **(a)** A fluorescence image of the single NV center over a scanning range of 1µm × 1 µm. **(b)** Fluorescence in dependence on x- and y-positions. The solid lines show the fits to gaussian functions. The increase of FWHM as compared to objective limit is attributed to the effect of resonator optical access hole.

The simple geometrical approach employed here can be extended for analysis of these parameters at any position within the optically accessible area. However, this approach neglects additional effects like reflections on the diamond-oil interface, which should be considered for a more precise analysis.

## S5. Experimental details

The experiments were performed on a home-built setup, whose schematic is shown as Fig. S9. Single NV centers in diamond can be optically addressed, initialized and detected with a confocal microscope. Here we used a diode-pumped solid state continuous wave laser with a wavelength of 532 nm (marked in green in the schematic) for the optical excitation. For pulsed experiments, we use an acousto-optical modulator (AOM) to generate pulses from the continuous wave laser beam or a pulsed diode laser. The microscope objective is fixed to the nanopositioning system that scans the sample in three dimensions. The fluorescence light (marked in red in the schematic) is also collected by this MO lens and passes through the dichroic mirror to the avalanche photodiode (APD) detector, while the scattered laser light is reflected by the dichroic mirror.

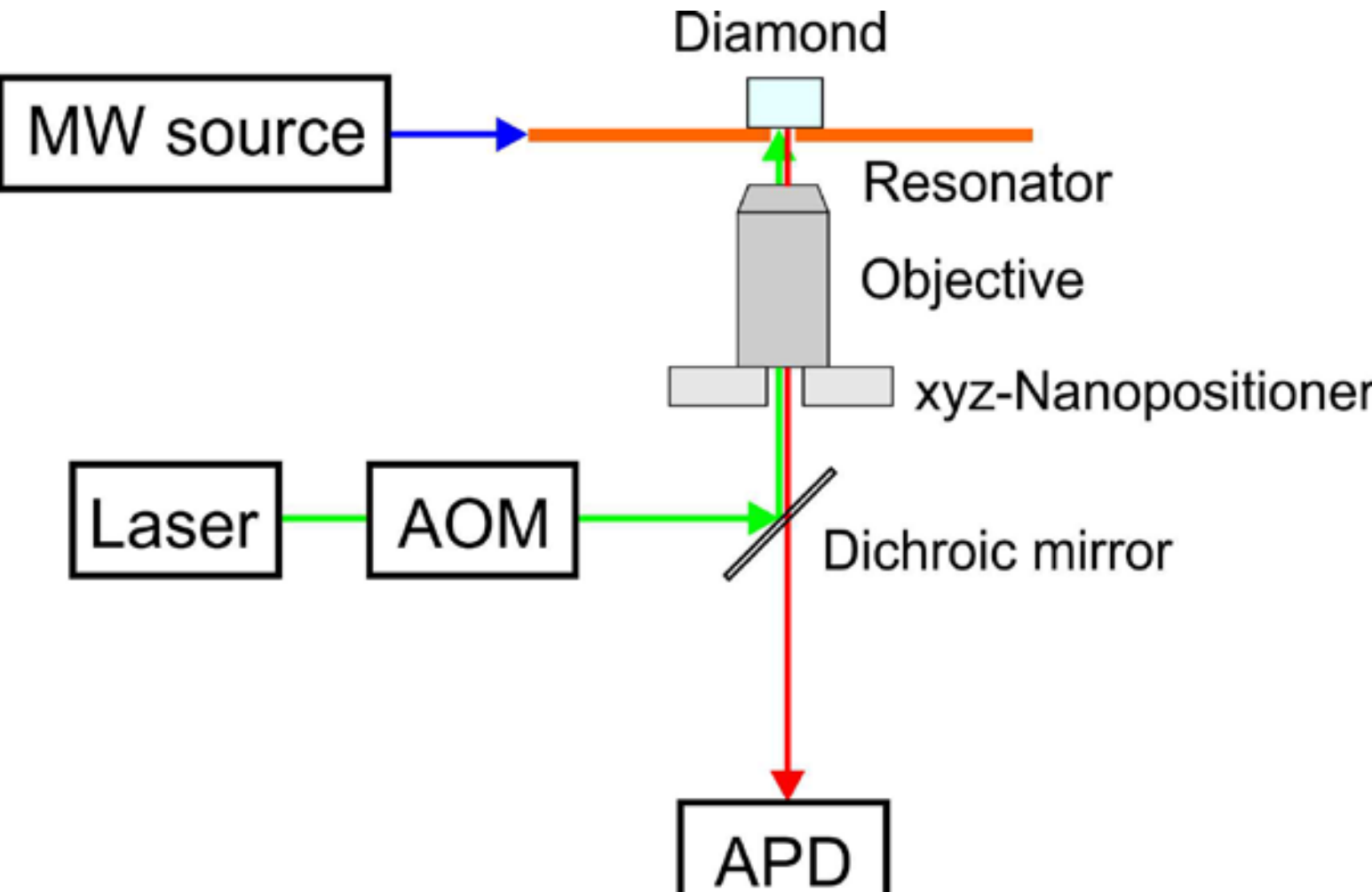

**Figure S10.** Schematic of the optical part of the setup. A confocal microscope is utilized for initializing and detecting single NV centers.

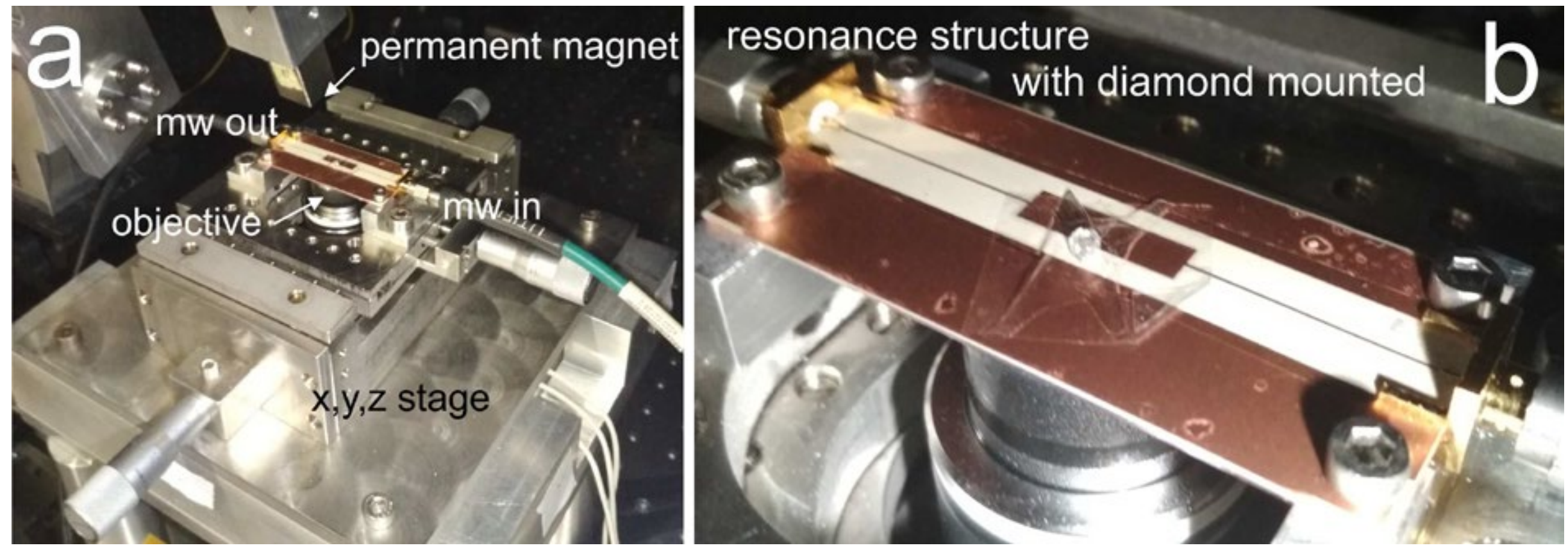

**Figure S11.** Photographs of the experimental setup.

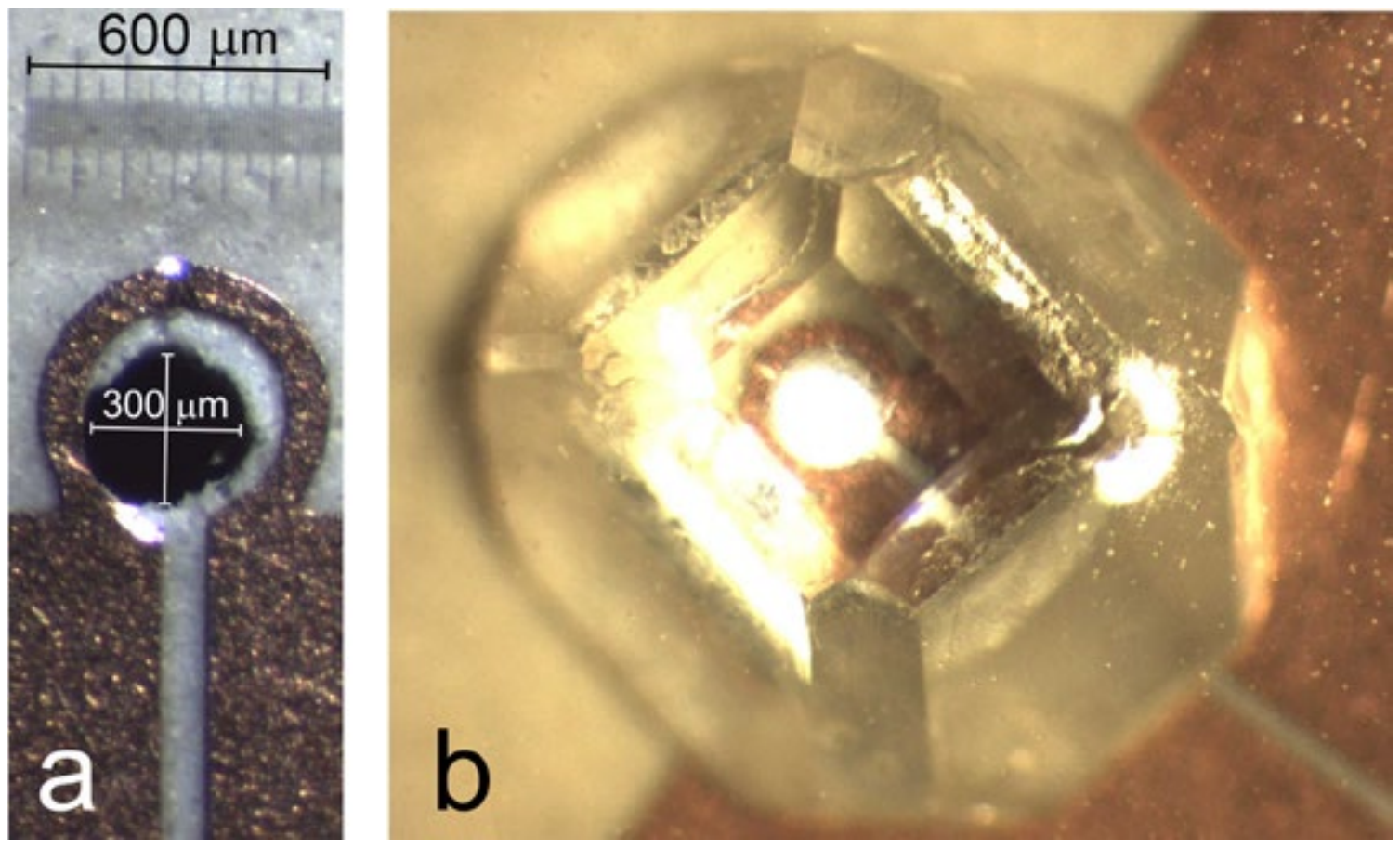

**Figure S12.** Photographs of **(a)** loop region of the resonator and **(b)** the resonator loop region with a diamond mounted.

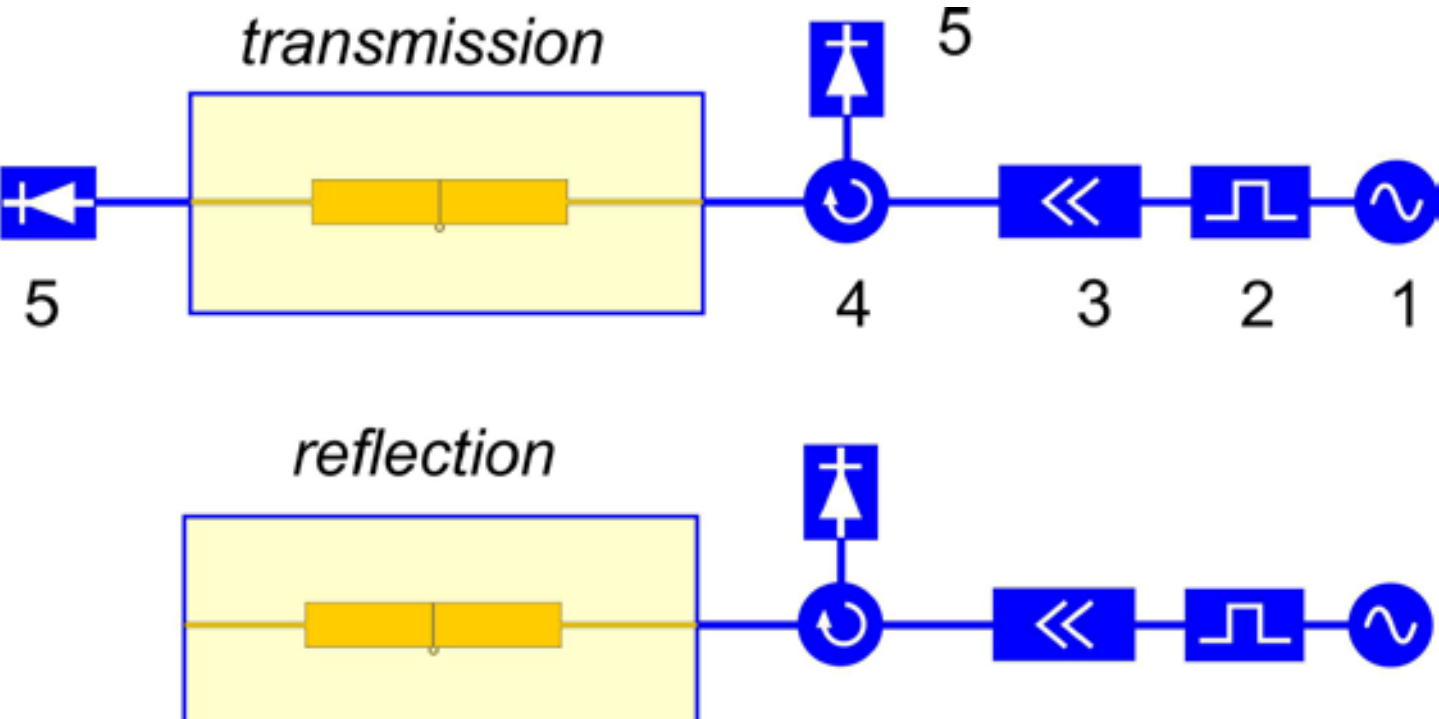

**Figure S13.** The principal microwave schemes for transmission and reflection modes. The components are: (1) cw microwave source; (2) PIN-diode modulator; (3) power mw amplifier; (4) circulator; (5) fast microwave diode detector for power and matching control.

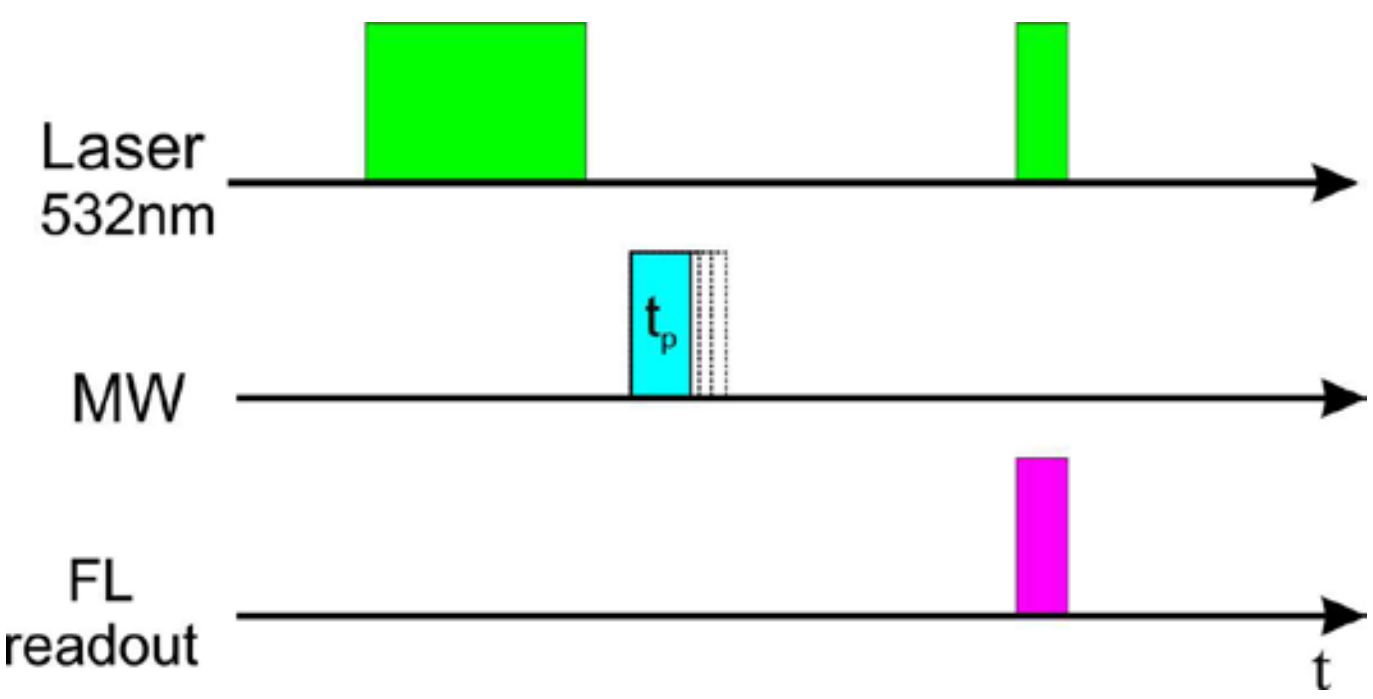

**Figure S14.** Pulse sequence for measuring Rabi oscillation. The first laser pulse of 5 μs initializes the NV into the ground state |g⟩. The MW pulse of variable length, $t_p$, flips the NV to $\cos(\pi f_R t_p)|g\rangle + \sin(\pi f_R t_p)|e\rangle$,where $|e\rangle$ denotes one excited state. The second laser pulse of 0.4 μs probes the population of ground state $|g\rangle$, proportional to the FL intensity. The pulse repetition rate was generally set to 6400 $s^{-1}$.

**S6 Frequency tuning capability**

In this part we consider the options for tuning the resonance frequency. This will be relevant for experiments in different magnetic fields or on other types of optically active spin centers.

The half-wave mode ($\lambda_g/2$) resonance frequency of a microstrip of length $L$ without current loop is given by the well-known expression

$$\nu_0 = \frac{c}{2L \cdot \sqrt{\varepsilon_{ff}}}$$

where $\varepsilon_{ff}$ is the effective dielectric constant of the microstrip. For a microstrip of width $W$, larger than the dielectric thickness $H$, it is given by

$$\varepsilon_{eff} = \frac{\varepsilon_r + 1}{2} + \frac{\varepsilon_r - 1}{2} \frac{1}{\sqrt{1 + 12\frac{H}{W}}}$$

The effective dielectric constant of $\varepsilon_{eff}$ =2.81 is calculated for microstrip with $H = 0.13$ mm, $W = 3$ mm and $\varepsilon_r = 3$ used in this work. Thus, for the fixed geometry of the resonator holder (see Fig 1, main text) the resonance frequency of the microstrip can be tuned from 1.6 GHz ($L$ = 56mm) to 18 GHz ($L$ = 5mm), see Fig. S15(a). The validity of the analytic approach is also valid for the full model as the resonance frequency from EM calculations (red line, Fig. S15(a)) agrees well with the analytical prediction (dashed line in Fig. S15(a)).

The blue line in Fig. S15(a) shows the resonance frequency of the microstrip resonator including the current loop in transmission mode. The additional gap and the current loop increase the intrinsic capacity and inductance of the microstrip. Thus, the resonance frequency is lower than for a microstrip of the same length. The resonance frequency can be tuned between 1.3 GHz ($L$ = 56mm) and 5.7 GHz ($L$ = 5mm). It is important to note that the conversion efficiency does not strongly depend on the resonance frequency, see Fig. 15(b). A slight efficiency improvement is observed at lower frequencies, accompanied by a reduction of the resonator bandwidth.

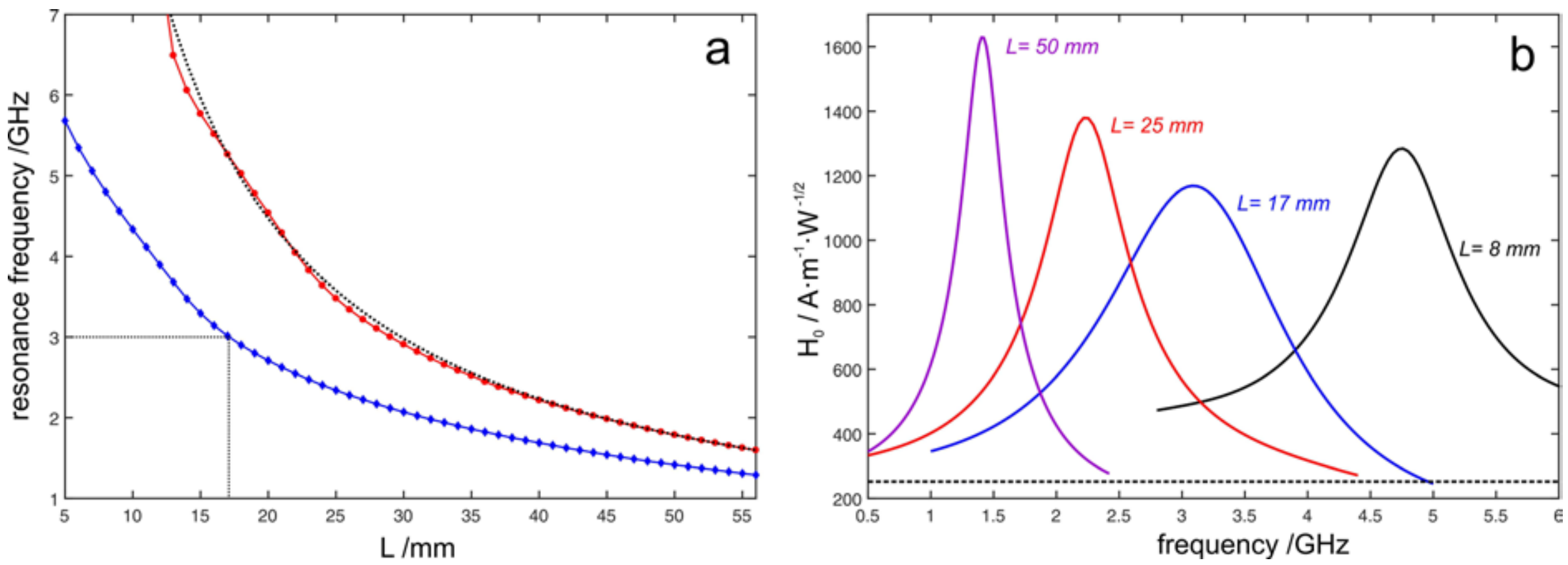

**Figure S15. (a)** Dependence of the resonance frequency on the length $L$, (diamonds, blue line) calculated for transmission mode. All other geometrical parameters are fixed as given in the caption of Figure 1 (main text). The red line with circles shows the resonance frequency of the microstrip without loop obtained from EM simulations for transmission mode. The black dashed line shows the analytically calculated resonance frequency of the microstrip. **(b)** Conversion efficiency calculated for resonators of selected lengths.

Thus, the resonator configuration described in this work can be used for experiments with microwave excitation over the frequency range of 1 GHz to 6 GHz. The range can be extended to lower frequencies by further increase of the resonator length or by increasing the dielectric constant of the substrate. Figure S16 shows the parameters of identical resonators with L = 50 mm calculated for ε = 3 (used in this work) and ε = 10. The ratio of the resonance frequencies 1.44 GHz/0.78 GHz = 1.85 agrees well with the ratio of the effective dielectric constants $\sqrt{9.15/2.81}$=1.8. The large reduction in the bandwidth (300 MHz vs. 110 MHz) can be improved by decreasing the width of the resonator, $W_R$, which would lead to an increase of the coupling coefficient (see section S1.3).

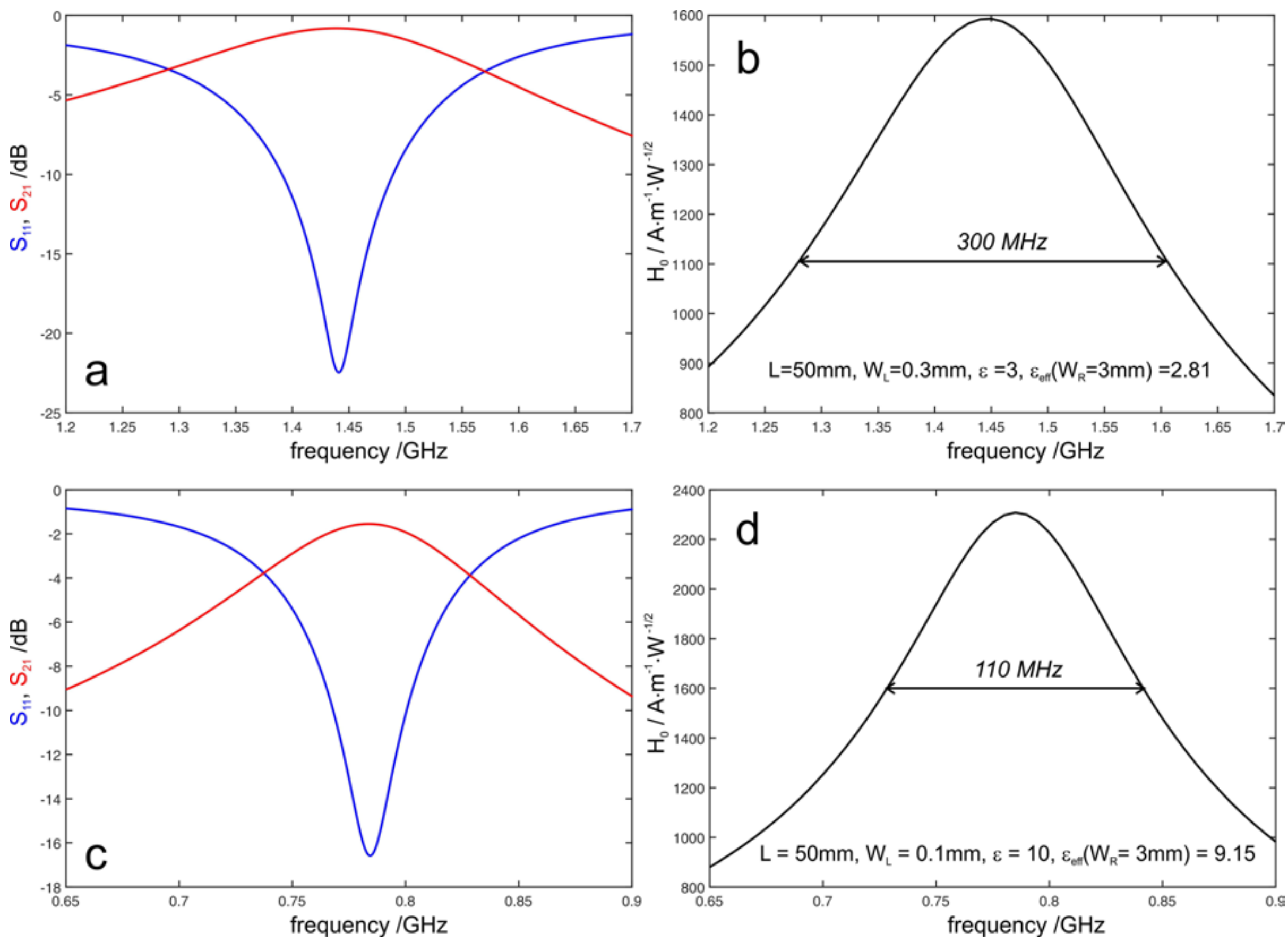

**Figure S16. (a,c)** S-parameters of identical transmission mode resonators with length $L = 50$ mm calculated for substrates having ε = 3 (a) and ε = 10 (c). All other geometrical parameters are fixed as given in the caption of Figure 1 (main text). **(c,d)** Corresponding conversion efficiencies. Resonator bandwidths and key parameters are indicated.

An increase of the resonance frequency above 6 GHz is difficult because substrates with dielectric constant < 2 are not available and a further decrease of the resonator length below $L \leq 2 \cdot W_R$ is not feasible, see Table S2. The resonator, however, can still be used at higher mw frequencies if higher resonance modes are considered. Figure S17 shows $S_{11}$ and efficiency parameters of the resonator (see Fig. 1 main text) extended to higher mw frequencies, where additional resonances are observed. Of particular interest is the resonance at about 11 GHz which corresponds to the $\sim\frac{3}{2}\lambda_g$ resonance mode of the microstrip. At this mode the conversion efficiency (600 A/m/$\sqrt{\text{W}}$) is reduced by factor 1.95 as compared to the main mode (1170 A/m/$\sqrt{\text{W}}$) but it is still about factor 2.5 higher compared to the 50 Ω microstrip with current loop (245 A/m/$\sqrt{\text{W}}$). We note here that for a microstip resonator without current loop in $\frac{3}{2}\lambda_g$ mode one would expect a reduction of the magnetic field amplitude in the center by a factor $\sqrt{3}$ compared to the fundamental $\frac{1}{2}\lambda_g$ mode. This is due to an additional resonance mode at 10.4 GHz, marked by * in Fig. S17. This mode generates only small magnetic fields at the position of the current loop and is therefore not useful. Thus, resonator designs operating at higher modes would require additional optimization steps in order to separate active from inactive modes.

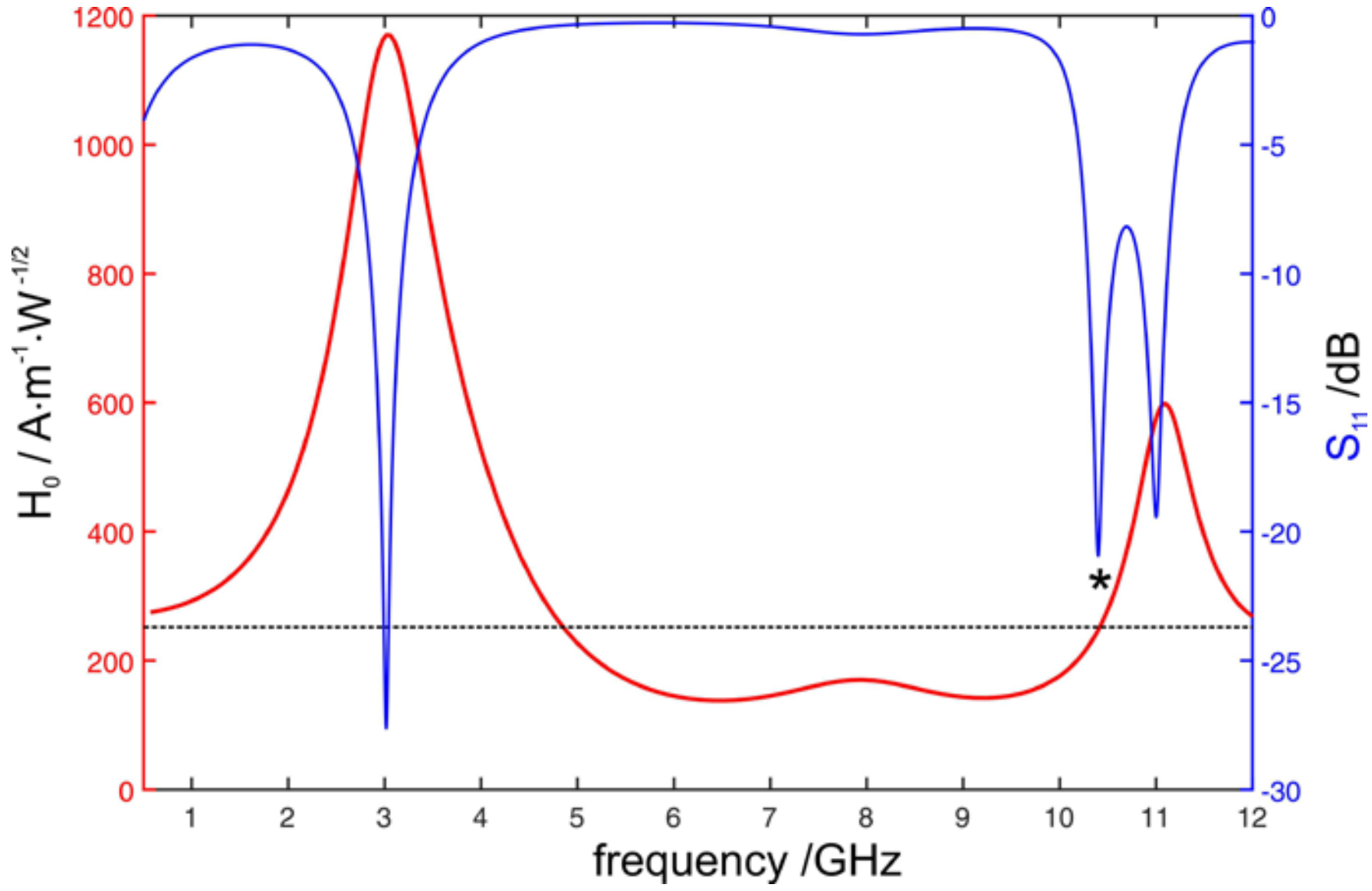

**Figure S17.** Simulated $S_{11}$ parameter and conversion efficiency of the transmission mode resonator described in Fig. 1 (main text) over a wider frequency range. The dashed line indicates the conversion efficiency of a 50 Ω microstrip line containing a current loop. The asterisk marks the frequency position of a magnetically inactive mode.